\documentclass[iop]{emulateapj}

\usepackage{amsmath}
\usepackage{color}
\usepackage[normalem]{ulem}
\usepackage[plainpages=false, colorlinks=true, anchorcolor=blue,
linkcolor=blue, citecolor=blue, bookmarks=false]{hyperref}

\usepackage{natbib}
\defcitealias{Dr11}{Dr11}
\defcitealias{KM09}{KM09}
\defcitealias{MQT10}{MQT10}

\newcommand{\p}{\partial}
\newcommand\pc{\;{\rm pc}}
\newcommand\second{\;{\rm s}}
\newcommand\yr{\;{\rm yr}}
\newcommand\Myr{\;{\rm Myr}}
\newcommand\cm{\;{\rm cm}}
\newcommand\gram{\;{\rm g}}
\newcommand\kms{\;{\rm km}\,{\rm s}^{-1}}
\newcommand\Msun{\;M_{\odot}}
\newcommand\Lsun{{\;L_\odot}}
\newcommand\Kel{\;{\rm K}}
\newcommand\kB{{\,k_{\rm B}}}
\newcommand\eV{{\,{\rm eV}}}

\newcommand\HII{\ion{H}{2} }
\newcommand\muH{\mu_{\rm H}}
\newcommand\alphaB{\alpha_{\rm B}}
\newcommand\fion{f_{\rm ion}}
\newcommand\fionch{f_{\rm ion,ch}}
\newcommand\Tion{T_{\rm i}}
\newcommand\sigmapi{\sigma_{\rm p.i.}}
\newcommand\sigmad{\sigma_{\rm d}}
\newcommand\Qiunit{Q_{\rm i,49}}
\newcommand\Qi{Q_{\rm i}}
\newcommand\Li{L_{\rm i}}
\newcommand\Ln{L_{\rm n}}
\newcommand\hnui{\langle h\nu \rangle_{\rm i}}
\newcommand\kappaIR{\kappa_{\rm IR}}
\newcommand\tauIR{\tau_{\rm IR}}

\newcommand\Rcl{R_{\rm cl}}
\newcommand\Mcl{M_{\rm cl}}
\newcommand\Mstar{M_{*}}
\newcommand\SFE{\varepsilon}
\newcommand\SFEm{\varepsilon_{\rm min}}
\newcommand\SFES{\varepsilon_{\rm S}}
\newcommand\Sigmacl{\Sigma_{\rm cl}}
\newcommand\krho{k_{\rho}}
\newcommand\avir{\alpha_{\rm vir}}
\newcommand\vturb{v_{\rm turb}}
\newcommand\vbind{v_{\rm bind}}

\newcommand\Ftot{F_{\rm tot}}
\newcommand\Fout{F_{\rm out}}
\newcommand\Fin{F_{\rm in}}
\newcommand\FIR{F_{\rm rad,IR}}

\newcommand\rsh{r_{\rm sh}}

\newcommand\vsh{v_{\rm sh}}
\newcommand\psh{p_{\rm sh}}
\newcommand\rIF{r_{\rm IF}}
\newcommand\rIFo{r_{\rm IF,0}}
\newcommand\rch{r_{\rm ch}}
\newcommand\Msh{M_{\rm sh}}
\newcommand\nedge{n_{\rm edge}}
\newcommand\nmean{n_{\rm mean}}
\newcommand\nrms{n_{\rm rms}}
\newcommand\nrmso{n_{\rm rms,0}}
\newcommand\taud{\tau_{\rm d}}
\newcommand\taudIF{\tau_{\rm d,IF}}
\newcommand\rc{r_{\rm c}}
\newcommand\nc{n_{\rm c}}
\newcommand\Qitom{\Xi}
\newcommand\Ltom{\Psi}
\newcommand\tff{t_{\rm ff}}

\begin{document}
\title{Disruption of Molecular Clouds by Expansion of Dusty \HII Regions} %

\shorttitle{Cloud Disruption by Expanding \HII Regions} %

\shortauthors{Kim et al.}

\author{Jeong-Gyu Kim\altaffilmark{1,2}, Woong-Tae
  Kim\altaffilmark{1}, \& Eve C.~Ostriker\altaffilmark{2}}

\affil{$^1$Center for the Exploration of the Origin of the Universe
  (CEOU), Astronomy Program, Department of Physics \& Astronomy,\\
  Seoul National University, Seoul 08826, Republic of Korea} %
\affil{$^2$Department of Astrophysical Sciences, Princeton University,
  Princeton, NJ 08544, USA} %
\email{jgkim@astro.snu.ac.kr, wkim@astro.snu.ac.kr, eco@astro.princeton.edu} %
\submitted{Accepted by \apj}

\begin{abstract}
Dynamical expansion of \HII regions around star clusters plays a key
role in dispersing the surrounding dense gas and therefore in limiting
the efficiency of star formation in molecular clouds. We use a
semi-analytic method and numerical simulations to explore expansion of
spherical dusty \HII regions and surrounding neutral shells and the
resulting cloud disruption. Our model for shell expansion adopts the
static solutions of \citet{Dr11} for dusty \HII regions and considers
the contact outward forces on the shell due to radiation and thermal
pressures as well as the inward gravity from the central star and the
shell itself. We show that the internal structure we adopt and the
shell evolution from the semi-analytic approach are in good agreement
with the results of numerical simulations. Strong radiation pressure
in the interior controls the shell expansion indirectly by enhancing
the density and pressure at the ionization front. We calculate the
minimum star formation efficiency $\SFEm$ required for cloud
disruption as a function of the cloud's total mass and mean surface
density. Within the adopted spherical geometry, we find that typical
giant molecular clouds in normal disk galaxies have
$\SFEm\lesssim10\%$, with comparable gas and radiation pressure
effects on shell expansion. Massive cluster-forming clumps require a
significantly higher efficiency of $\SFEm\gtrsim50\%$ for disruption,
produced mainly by radiation-driven expansion. The disruption time is
typically of the order of a free-fall timescale, suggesting that the
cloud disruption occurs rapidly once a sufficiently luminous \HII
region is formed. We also discuss limitations of the spherical
idealization.
\end{abstract}

\keywords{galaxies: star clusters --- galaxies: star formation ---
  \HII regions --- ISM: clouds --- ISM:kinematics and dynamics ---
  stars: formation}

\section{Introduction}
Giant molecular clouds (GMCs) are the sites of star formation in
galaxies. They are highly structured, consisting of hierarchy of
clumps, filaments, and sheets resulting from shock interactions in
supersonic turbulence \citep{elm04,and14}.  Stars predominantly form
in groups (later becoming OB associations or clusters) within dense,
gravitationally-bound clumps inside GMCs \citep{lad03}. Newborn star
clusters have a profound influence on the surrounding interstellar
medium (ISM) via protostellar outflows, stellar winds, ionizing
radiation, and supernova explosions, which are collectively referred
to as stellar feedback. The question of how each feedback process
affects formation, evolution, and dispersal of their natal clouds is
an active and contentious area of research (see \citealt{dob14} and
\citealt{kru14} for recent reviews).

An important unsolved problem in star formation theory is what
determines net star formation efficiency $\SFE$ of a cloud, defined as
the fraction of the cloud's mass that is turned into stars over its
lifetime. GMCs are known to be inefficient in converting gas into
stars. Over their lifetime, individual GMCs in the Milky Way appear to
turn only a few to several percent of their mass into stars
\citep{mye86,wil97,car00,eva09,ken12,gar14}. Furthermore, observations
of Galactic infrared dark clouds, nearby galaxies, and high-redshift
star-forming galaxies all point to a conclusion that the depletion
time of molecular gas is more than an order of magnitude longer than
the internal dynamical timescale
\citep[e.g.,][]{kru07b,gen10,kru12a,ler13}.

Observations and theoretical arguments indicate that star formation
efficiency tends to be higher within high-density environments
\citep{elm97,mck07}. For example, the estimated star formation
efficiencies of cluster-forming clumps for low-mass clusters in the
solar neighborhood is $\sim 0.1$--$0.3$, higher than the few percent
efficiency of entire GMCs \citep{lad03}. The Orion Nebula Cluster,
which has been forming stars for several dynamical times
\citep{tan06}, appears to have stellar fraction $\sim 50\%$
\citep{dar14}.  Observations of massive, dense clouds in dwarf
starburst galaxies containing nascent super star clusters indicate
efficiencies $> 50 \%$ \citep{mei02,tur15}. Gas expulsion from
protoclusters is also important in the context of disruption or
survival of a stellar cluster
\citep[e.g.,][]{hil80,elm83,goo97,lad03,ban15}. If star formation
efficiency is sufficiently high, a star cluster may remain
gravitationally bound and thus become long-lived. If star formation
efficiency varies with the mass of protocluster, it is likely to leave
an imprint on the shape of the cluster mass function distinct from the
cloud mass function \citep[e.g.,][]{ash01,kro02,fal10}.

A number of theoretical studies have proposed that \HII regions may be
the primary means of controlling the star formation efficiency within
a cluster's birth cloud \citep[e.g.,][]{whi79, wil97, mat02, kru06,
  gol11, dal12, dal13a}. A newborn cluster of stars embedded in a cloud
emits abundant ultraviolet (UV) photons, creating an ionization front
that separates the fully ionized gas close to the cluster from
surrounding neutral gas. Thermal balance between heating by
photoionization and cooling by line emission keeps the temperature of
the ionized gas roughly at $\sim10^4\Kel$
\citep[e.g.,][]{ost89,dr11}. In the absence of radiation pressure (see
below) and considering confined rather than blister-type \HII regions,
the density of ionized gas is relatively uniform.  Expansion of the
ionized gas due to its high thermal pressure (initially $\sim 10^3$
times ambient levels) drives a shock wave ahead of the ionization
front.  An expanding shell of dense gas between the ionization and
shock fronts is created, incorporating the ambient neutral gas as the
shock sweeps outward. Cloud disruption by shell expansion and/or
associated photoevaporation create hostile conditions for further star
formation. \citet{whi79} and \citet{fra94} found that photoevaporation
by massive stars born near the cloud boundary can limit $\SFE$ to
$\sim5\%$ in a typical molecular cloud. \citet{wil97} and
\citet{mat02} found that clouds convert on average $\sim10\%$ of their
mass into stars before destruction by photoevaporation. Using the
time-dependent virial theorem, \cite{kru06} found that both mass
ejection by photoevaporation and momentum injection by expanding \HII
regions limit the net star formation efficiency of GMCs to
$\sim5$--$10\%$ before disruption.

The classical picture of an embedded \HII region described above does
not account for the effects of radiation pressure on dust grains,
which are efficient at absorbing UV photons. If dust is tightly
coupled to gas through mutual collisions, radiation forces exerted on
the former are readily transmitted to the latter. Not only does dust
reduce the size of the ionized zone \citep{pet72}, but it can also
produce a central ``hole'' near the cluster by the action of radiation
pressure \citep{mat67, art04}. Recently, \citet[][hereafter
  Dr11]{Dr11} obtained families of similarity solutions for the
internal structure of dusty \HII regions in static force balance. He
found that radiation pressure is important for dense and luminous \HII
regions, forming a central cavity surrounded by an over-dense ionized
shell just inside the ionization front. \citet[][hereafter KM09]{KM09}
showed that while expansion of \HII regions excited by a small number
of massive stars is well described by the gas-pressure driven
classical model \citep{spi78}, the dynamics of \HII regions around
massive star clusters is dominated by radiation pressure. They
presented an analytic formula for the shell evolution driven by the
combination of gas and radiation pressures, finding that radiation
pressure is more important during the early phase of the expansion,
while the late stage is governed by a gas pressure force that
increases with the shell radius.

Since gas expulsion by thermally driven expanding \HII regions does
not occur efficiently for clouds with high escape velocities ($\gtrsim
10 \kms$) \citep[e.g.,][]{mat02, kru06, dal12, dal13a}, radiation
pressure has been considered the most promising mechanism for
disruption of massive clouds (e.g., \citealt{sco01};
\citetalias{KM09}; \citealt{fal10, MQT10}). In particular,
\citet{fal10} considered the potential for disruption of molecular
clouds by various feedback mechanisms including protostellar outflows,
photoionization, and supernova explosions, and concluded that
radiation pressure may dominate momentum injection in dense and
massive protoclusters.  Utilizing the analytic solutions of
\citetalias{KM09} for shell expansion, they argued that the minimum
star formation efficiency, $\SFEm$, required for cloud disruption
primarily depends on the mean cloud surface density. With $\SFEm$
independent of the mass, this would explain the observed similarity
between shapes of mass functions of molecular clouds and young star
clusters. \citet[][hereafter MQT10]{MQT10} analytically examined
disruption of massive GMCs for sample systems representative of
various galactic environments, estimating effects of stellar feedback
from protostellar jets, shocked stellar winds, thermal pressure of
photoionized gas, radiation pressure, and the total gravity from stars
and shell.  They found that star formation efficiency for massive GMCs
in the Milky Way is only a few percent, while clouds in starburst
galaxies and star-forming giant clumps in high-redshift galaxies
require $\sim 20$--$40 \%$ of efficiency for disruption.  In the part
of parameter space they explored, direct and/or dust-reprocessed
radiation pressure dominates ionized-gas pressure in driving shell
expansion.

While the previous analytic works mentioned above are informative in
understanding the effects of radiation pressure on the shell expansion
and related cloud disruption, they are not without limitations. The
minimum star formation efficiency derived by \citet{fal10} does not
allow for the effect of gas pressure and inward gravity that may be
important in the late stage of the shell expansion. \citetalias{MQT10}
applied their expansion model to only a few representative cases, so
that the general dependence of $\SFE$ on the cloud mass and surface
density has yet to be explored. In addition, \citetalias{KM09} and
\citetalias{MQT10} set the radiation force on the shell equal to
$L/c$, where $L$ and $c$ refer to the luminosity of the central source
and the speed of light, respectively, assuming that all dust inside
the \HII region is pushed out to the shell. They also assumed the
thermal pressure acting at the inner edge of the shell is equal to the
mean thermal pressure in the ionized region. Since the static
solutions of \citetalias{Dr11} indicate the ionized gas inside a dusty
\HII region is strongly stratified (at high luminosity) and absorbs
radiation from the central source, the assumption of the unattenuated
radiation up to the shell should be checked. All of the above also
adopt the assumption of spherical symmetry and a source luminosity
that is constant in time.

The numerical work of \citet{dal12,dal13a} is not limited by symmetry
idealizations, allowing for fully turbulent gas dynamics and
self-consistent collapse to create sources of ionizing
radiation. However, these works do not include effects of radiation
pressure, and consider only a limited parameter space. Very recently,
\citet{ras15} have conducted a set of numerical radiation hydrodynamic
(RHD) simulations focusing exclusively on the effects of
(non-ionizing) UV in turbulent clouds with surface densities in the
range $10$--$300 \Msun \pc^{-2}$. The \citeauthor{ras15} simulations
showed that turbulent compressions of gas can raise the value of
$\SFE$ by a factor of $\sim$5 for typical Milky Way GMC parameters,
because strong radiation forces are required for dispersal of dense,
shock-compressed filamentary structures. \citet{ski15} used numerical
RHD simulations to consider the complementary regime of extremely high
surface density clouds, evaluating the ability of radiation forces
from dust-reprocessed infrared (IR) to disrupt clouds. This work
showed that IR is effective only if $\kappaIR > 15 \cm^2 \gram^{-1}$,
with $\kappaIR$ being the gas opacity to dust-reprocessed IR
radiation, and even in this case the predicted efficiency is $\sim
50\%$, which may explain observations of nascent super star clusters
\citep{tur15}.

In this paper, we use a simple semi-analytic model as well as
numerical simulations to investigate expansion of dusty \HII regions
and its effect on disruption of star-forming clouds across a variety
of length and mass scales. This work improves upon previous analytic
works in several ways. First, shells in our model expand due to both
radiation and thermal pressures explicitly, extending \citet{fal10},
which considered solely radiation pressure. We also include the inward
force due to gravity of the stars and shell, which extends the
non-gravitating models of \citetalias{KM09}. Second, we adopt the
static solutions of \citetalias{Dr11} for non-uniform internal
structure of \HII regions. This allows us not only to accurately
evaluate the contact forces on the shell arising from thermal and
radiation pressures, but also to compare them with the effective
forces adopted in the previous studies \citepalias[e.g.,][]{KM09,
  MQT10}.  Third, we have also run direct numerical simulation to
check the validity of the solutions of \citetalias{Dr11} for
representing the interior structure in an expanding \HII region and
also to confirm our semi-analytic shell expansion solutions. Fourth,
we conduct a systematic parameter survey in the full space of cloud
mass and surface density, evaluating at each ($\Mcl$, $\Sigmacl$) the
minimum efficiency $\SFEm$ required for disruption, the relative
importance of radiation and ionized-gas pressures, and the timescale
of cloud disruption. We also consider the effects of the
mass-dependent light-to-mass ratio, the density distribution of the
background medium, and the dust-reprocessed radiation. Our model does
not allow for potentially important effects of stellar winds, which we
briefly discuss in Section~\ref{s:dis}. The chief idealization of our
study is the adoption of spherical symmetry. While real clouds are not
symmetric, our results provide a guide and baseline for future work
that will relax this restriction.

The rest of this paper is organized as follows. In
Section~\ref{s:dr11}, we briefly summarize the solutions of
\citetalias{Dr11} for dusty \HII regions in static equilibrium. In
Section~\ref{s:exp}, we describe our semi-analytic model for shell
expansion, and evaluate the contact forces on the shell in comparison
with the effective forces adopted by other authors. We also run
numerical simulations for expanding \HII regions, and compare the
results with those of the semi-analytic model. In Section~\ref{s:sfe},
we calculate the minimum efficiency of star formation required for
cloud disruption. We also present analytic expressions for $\SFEm$ in
the radiation or gas pressure driven limits. Finally, we summarize our
results and discuss their astrophysical implications in
Section~\ref{s:sum}.

\section{Internal Structure of Dusty \HII Regions}\label{s:dr11}

\citetalias{Dr11} studied the effect of radiation pressure on the
internal structure of static, dusty \HII regions, finding that
radiation pressure acting on gas and dust gives rise to a non-uniform
density profile with density (and gas pressure) increasing outward. In
this section, we revisit \citetalias{Dr11} to obtain the parametric
dependence of various \HII region properties on the strength of
ionizing radiation. This information will be used for our dynamical
models in Section \ref{s:exp}.

Let us consider a central point source with luminosity $L=\Li +\Ln$
embedded in a cloud, where $\Li$ and $\Ln$ refer to the luminosities
of hydrogen ionizing and non-ionizing photons, respectively.  The
number of ionizing photons per unit time emitted from the source is
$\Qi = \Li/\hnui$, where $\hnui$ is the mean photon energy above the
Lyman limit. For simplicity, we ignore the effect of He and assume
that the ionized gas has a constant temperature $\Tion=10^4 \Kel$
under photoionization equilibrium. Dust grains absorb both ionizing
and non-ionizing radiation, with a constant absorption cross-section
per hydrogen nucleus of $\sigmad= 10^{-21} \cm^2\,\rm{H}^{-1}$
\citepalias{Dr11}. The outward photon momentum absorbed by the dust is
transferred to the gas via thermal and Coulomb collisions, resulting
in a non-uniform gas density profile $n(r)$. Let $\fion$ denote the
fraction of photons absorbed by the gas. Assuming ``Case B''
recombination, the radius of the Str{\"o}mgren sphere centered at the
source is then given by
\begin{equation}\label{e:rIF}
  \rIF \equiv \left( \dfrac{3\fion\Qi}{4\pi\alphaB\nrms^2}
  \right)^{1/3}\,,
\end{equation}
where $\alphaB=2.59 \times 10^{-13} \left(T/10^4 \Kel\right)^{-0.7}
\cm^{3}\second^{-1}$ is the effective recombination coefficient
\citep{ost89} and
\begin{equation}
  \nrms \equiv \left(\frac{3}{\rIF^3}\int_0^{\rIF} n^2(r) r^2 dr\right)^{1/2}
\end{equation}
is the root-mean-square number density within the ionized region.

The luminosity $L(r)$ at radius $r$ inside the Str{\"o}mgren sphere is
given by $L(r) = \Li \phi(r) + \Ln e^{-\taud}$, where $\phi(r)$ is the
dimensionless quantity that describes attenuation of ionizing photons
and $\taud = \int_{0}^{r} n\sigmad dr$ is the dust optical depth of
non-ionizing photons. The functions $\phi(r)$ and $\taud(r)$ ought to
satisfy
\begin{equation}\label{e:dr-1}
  \dfrac{d\phi}{dr} = - n\sigmad\phi - 4\pi\alphaB n^2r^2/\Qi\,,
\end{equation}
\begin{equation}\label{e:dr-2}
  \dfrac{d\taud}{dr} = n\sigmad\,.
\end{equation}
The gas/dust mixture (assumed to be collisionally coupled) in the \HII
region is subject to both thermal pressure $P_{\rm
  thm}(r)=2n(r)\kB\Tion$ and radiation pressure $P_{\rm
  rad}(r)=L(r)/(4\pi r^2c)$. The condition of static force balance can
thus be written as
\begin{equation}\label{e:dr-3}
  \dfrac{1}{4\pi r^2 c}\dfrac{d}{dr}\left(\Li\phi + \Ln
  e^{-\taud}\right) + 2\kB \Tion \dfrac{dn}{dr} = 0\,.
\end{equation}

Equations~\eqref{e:dr-1}--\eqref{e:dr-3} can be solved simultaneously
for $n(r)$, $\phi(r)$, and $\taud(r)$ subject to the boundary
conditions $\taud(0)=0$, $\phi(0)=1$, and $\phi(\rIF)=0$.
\citetalias{Dr11} showed that there exists a family of similarity
solutions that are completely specified by three parameters: $\beta
\equiv \Ln/\Li$, $\gamma \equiv 2c\kB\Tion\sigmad/(\alphaB\hnui)$, and
$\Qi\nrms$, characterizing the (inverse of) importance of ionizing
photons, the dust content, and the radiation pressure, respectively.
Throughout this paper, we take $\beta=1.5$, $\hnui=18\eV$, and $\gamma
= 11.1$ as standard values, appropriate for \HII regions formed around
massive star clusters (see Appendix~\ref{s:ltom}).

\begin{figure}
\epsscale{1.15}\plotone{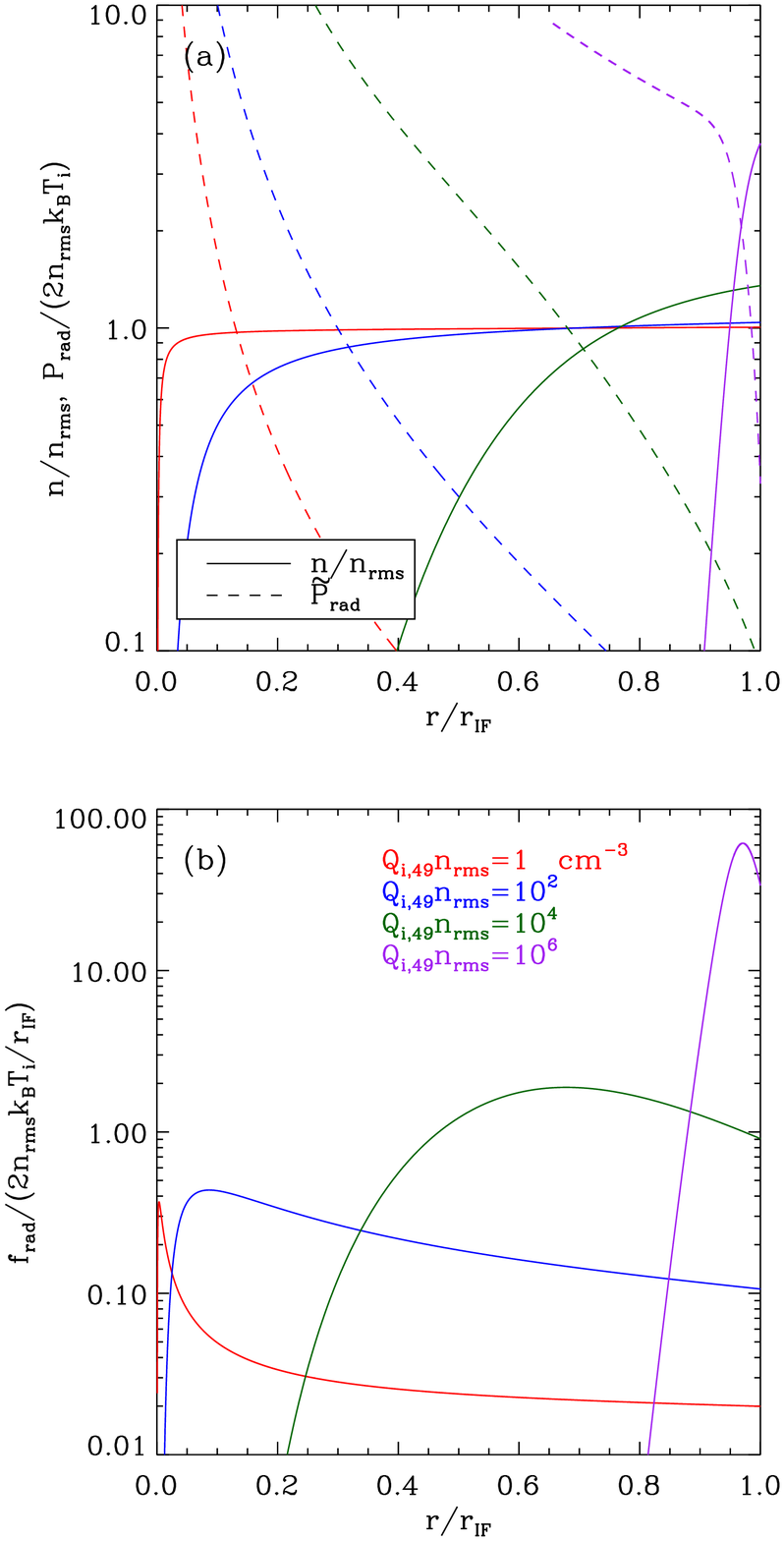} \caption{Radial distributions of (a)
  the normalized density $n/\nrms$ (proportional to gas pressure,
  solid lines) and the normalized radiation pressure $\tilde P_{\rm
    rad}= P_{\rm rad}/(2\nrms\kB\Tion)$ (dashed lines), and (b) the
  normalized radiation force per unit volume $\tilde f_{\rm rad} =
  f_{\rm rad}/(2\nrms\kB\Tion/\rIF)$ (equal to gas pressure force per
  unit volume in magnitude and opposite in direction) for
  $\Qiunit\nrms=1, 10^2, 10^4, 10^6\;\cm^{-3}$.}\label{f:dr11}
\end{figure}

Figure~\ref{f:dr11} plots the radial profiles of the normalized gas
density and radiation pressure in the upper panel, and the radiation
force per unit volume $f_{\rm rad} = -r^{-2}d(r^2P_{\rm rad})/dr$ in
the lower panel, for $\Qiunit\nrms =1, 10^2, 10^4, 10^6 \cm^{-3}$,
where $\Qiunit\equiv \Qi/(10^{49}\second^{-1})$.  For $\Qiunit\nrms
\lesssim 10^4 \cm^{-3}$, the internal structure is dominated by gas
pressure, except in the very central regions where $P_{\rm rad}>P_{\rm
  thm}$, resulting in an almost flat density distribution with $n
\approx \nrms$. In this case, the dust optical depth up to $r=\rIF$ is
less than unity, indicating that a large fraction of non-ionizing
radiation survives absorption by dust. For $\Qiunit\nrms \gtrsim 10^4
\cm^{-3}$, on the other hand, radiation pressure is crucial in
controlling the density structure. Gas close to the center is pushed
out to large radii, forming a central cavity and an outer ionized
shell, making the density at the ionization front $\nedge\equiv
n(\rIF)$ larger than $\nrms$.\footnote{Dust-deficient \HII regions
  with $\gamma < 1$ do not possess a cavity (\citetalias{Dr11},
  \citealt{yeh12}).} Note that although $P_{\rm rad} \gg P_{\rm thm}$
in the cavity, the gas density there is too small to attenuate
radiation significantly. Consequently, the radiation force exerted on
the gas is almost negligible in the central cavity and rises only in
the outer parts of the \HII region, as Figure~\ref{f:dr11}(b)
illustrates. In fact, Equation (\ref{e:dr-3}) shows that $f_{\rm rad}$
achieves its maximum at the position of the steepest density gradient
\citep[see also][]{lop11,lop14}. Thus, radiation strongly modifies the
interior of the \HII regions such that both gas and radiation forces
are applied primarily in the outer portion.

\begin{figure}
\epsscale{1.2}\plotone{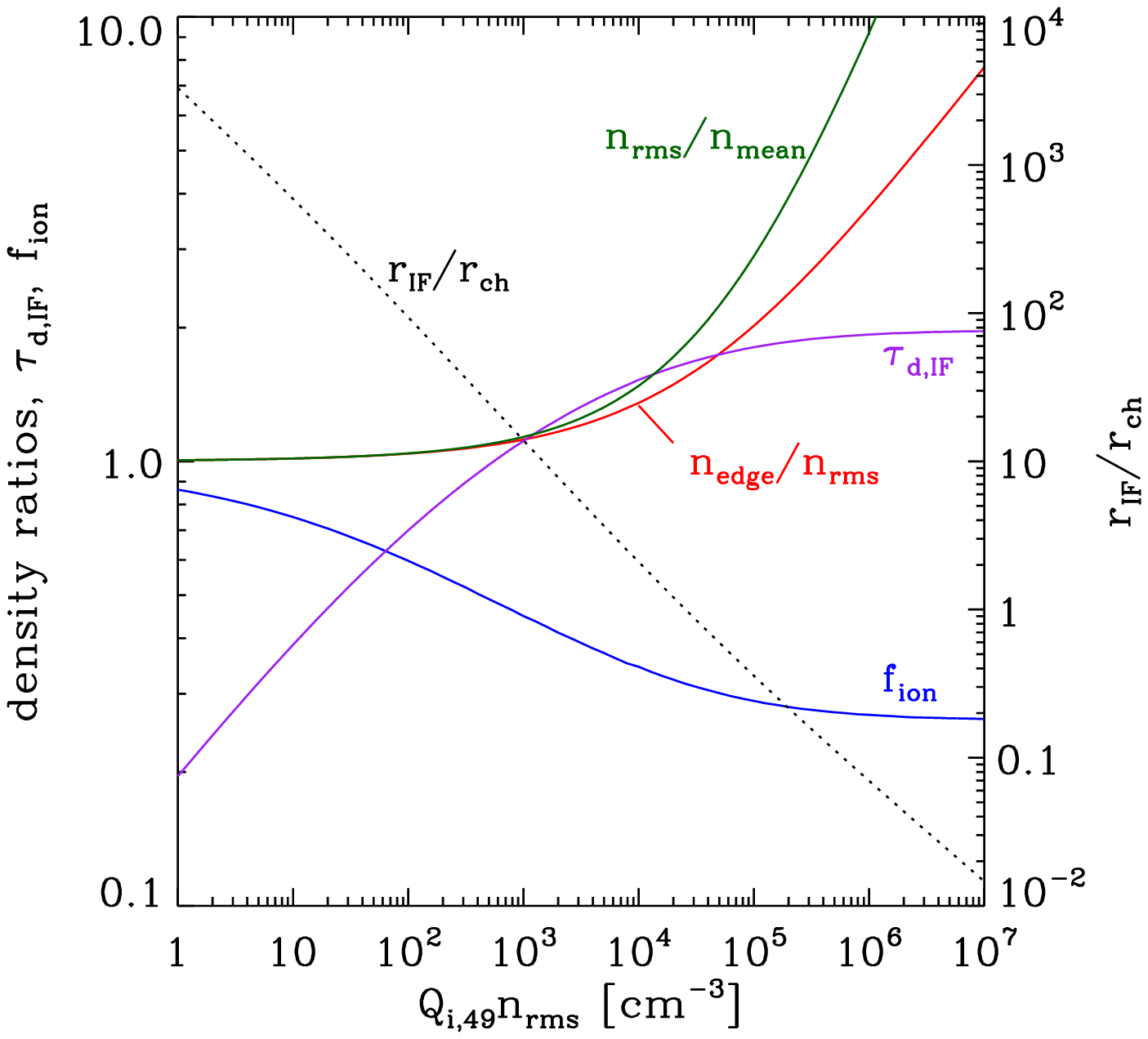} \caption{Dependence on
  $\Qiunit\nrms$ of the ratios of the edge-to-rms densities
  $\nedge/\nrms$ (red), the rms-to-mean densities $\nrms/\nmean$
  (green), the dust optical depth to the ionization front $\taudIF$
  (purple), and the fraction $\fion$ (blue) of ionizing photons
  attenuated by photoionization prior to reaching the shell. The black
  dotted line shows the relationship between $\Qiunit\nrms$ and the
  ratio $\rIF/\rch$ of the ionization front radius to the
  characteristic radius (see Equations~\eqref{e:rch} and
  \eqref{e:rIFrch}).}\label{f:dr11_param}
\end{figure}

Figure~\ref{f:dr11_param} plots the dependence on $\Qiunit\nrms$ of
$\taudIF\equiv\taud(\rIF)$, $\fion$, $\nedge/\nrms$, and
$\nrms/\nmean$, where $\nmean=3\int_0^{\rIF} n(r) r^2dr/\rIF^3$ is the
mean density. Note that $\taudIF$ and $\fion$ saturate to finite
values ($\taudIF=1.97$ and $\fion=0.26$) as $\Qiunit\nrms$
increases. This is consistent with the results of \citet{yeh12} who
showed that Equations \eqref{e:dr-1}--\eqref{e:dr-3} yield
\begin{equation}\label{e:taudlim}
  \gamma(\beta + 1)e^{-\frac{\gamma - 1}{\gamma}\taudIF} =
  \beta (\gamma -1 ) e^{-\taudIF} + \beta + 1\,,
\end{equation}
\begin{equation}\label{e:fionlim}
  \fion = \dfrac{\beta + 1}{\gamma -1}\left[ \taudIF -
    \dfrac{\gamma}{\gamma - 1} \left( 1 -
    e^{-\frac{\gamma-1}{\gamma}\taudIF} \right) \right]\,,
\end{equation}
\begin{equation}\label{e:nedgelim}
  \frac{\nedge}{\nrms} = \dfrac{1 + \beta(1 -
    e^{-\taudIF})}{3\fion\gamma}\nrms\rIF\sigmad
    \propto (\nrms \Qi)^{1/3}\,,
\end{equation}
in the limit of $\Qi\nrms \rightarrow \infty$. The saturation of
$\taudIF$ and $\fion$ occurs because the radiation-induced ionized
shell near $r=\rIF$ has such large density that neutral hydrogen
produced by recombinations can compete with dust in absorbing ionizing
photons. This is in contrast to the classical case of uniform-density
\HII regions without radiation pressure, for which $\taudIF
\rightarrow \infty$, $\fion \rightarrow 0$ as $\Qi\nrms \rightarrow
\infty$ \citep{pet72}. Since the ionized shell has a saturated column
$\int_0^{\rIF} n dr = \taudIF/\sigmad$, the gas mass $M_{\rm ion}$
within a dusty \HII region in this limit depends on $\rIF$ as $M_{\rm
  ion} \propto \rIF^{2}$, as opposed to classical \HII regions for
which $\nmean \approx \nrms$ and $M_{\rm ion} \propto \nmean\rIF^3
\propto \rIF^{3/2}$ for given $\Qi$.

\section{Expansion of Dusty \HII Regions}\label{s:exp}

The high pressure in the interior of an \HII region compared to the
ambient levels leads to radial expansion. \citet{spi78} provides
approximate solutions for this expansion in the case where just gas
pressure is included, under the assumption that ambient gas is swept
up into a thin shell surrounding the ionized interior (see also
\citealt{fra90,shu92}). In this section, we explore expansion of
spherically-symmetric, dusty \HII regions driven by both thermal and
radiation pressures, making a thin-shell approximation. Here we ignore
pressure resulting from diffuse recombination radiation, which is
negligible in comparison to gas pressure (e.g., \citealt{hen98};
\citetalias{KM09,Dr11}). We also assume that the shell is optically
thin to IR photons emitted by dust grains, the effect of which will be
examined in Section~\ref{s:krhoIR}. Expansion models by
\citetalias{KM09} and \citetalias{MQT10} adopted the assumption that
all of the radiation is absorbed only by the shell, while ionizing
radiation produces a uniform pressure, uniform density interior. Here,
we relax these assumptions and calculate the direct outward forces
based on the \citetalias{Dr11} solutions.

\subsection{Model}

We consider a spherical neutral cloud with density distribution
\begin{equation}\label{e:iniden}
 n(r) =
 \begin{cases}
  \nc, &\text{for $r/\rc < 1$},\\ \nc(r/\rc)^{-\krho}, &\text{for
    $r/\rc \geq 1$},
 \end{cases}
\end{equation}
where $n_c$ and $r_c$ are the density and radius of a flat core,
respectively, and $\krho$ is an index of a surrounding power-law
envelope.\footnote{While GMCs do not appear to possess density
  stratification on a global scale \citep{mck07}, the density
  distribution of clumps within them is well described by a power-law
  profile including a central core \citep[see e.g.,][]{fra90,hil98}.}
A star cluster with total mass $\Mstar$, total luminosity $L$, and
total ionizing power $\Qi$ is born instantly at the cloud
center. Copious energetic photons emitted by the cluster begin to
ionize the surrounding medium, causing the ionization front to advance
to the initial Str\"omgren radius, $\rIFo$, within a few recombination
timescales ($\sim 10^3\yr$ for $n\sim10^2 \cm^{-3}$). The
overpressured, ionized gas creates a shock front that moves radially
outward, sweeping up the ambient neutral medium into a dense shell. We
assume that the \HII region remains in internal quasi-static
equilibrium throughout its dynamical expansion, which is reasonable
since the sound-crossing time over the ionized region is sufficiently
small compared to the expansion timescale. We further assume that all
of the swept-up gas resides in a thin shell of mass $\Msh$ located at
$\rsh$. Since we are interested in the evolution of the shell well
after the formation phase, we may take $\rc \ll \rIFo$ in describing
shell expansion in a power-law envelope.

The momentum equation for the shell is
\begin{equation}\label{e:shell}
  \dfrac{d (\Msh\vsh)}{dt} = \Ftot = \Fout - \Fin\,,
\end{equation}
where $\vsh = d\rsh/dt$ is the expansion velocity of the shell, and
$\Fout$ and $\Fin$ denote the outward and inward forces acting on the
shell, respectively. The inward force is due to the cluster gravity and
shell self-gravity, given by
\begin{align}\label{e:Fgrav}
  \Fin = \dfrac{G \Msh(\Mstar + \Msh/2)}{\rsh^2}\,,
\end{align}
so that the cluster gravity and gaseous self-gravity depend on $\rsh$
as $\rsh^{1-\krho}$ and $\rsh^{4 - 2\krho}$, respectively. The
radiation and thermal pressures give rise to the total outward force
\begin{equation}\label{e:fout}
  \Fout = \dfrac{\Ln }{c} e^{-\taudIF} +
  8\pi\kB\Tion\nedge\rsh^2\,.
\end{equation}
Note that $\Fout$ is the {\it contact} force acting on the surface
immediately outside the ionization front. Only non-ionizing photons
reach the neutral shell and exert the outward force, if the shell is
optically thick to them.\footnote{The condition that the dusty shell
  should be opaque to UV photons can be expressed as $\rsh \lesssim
  (\kappa_{\rm UV}\Msh/4\pi)^{1/2} = 8.4 \pc (\Msh/10^4 \Msun)^{1/2}$,
  where $\kappa_{\rm UV} = \sigmad/\muH = 4.3 \times 10^2 \cm^2
  \gram^{-1}$ \citep{tho15}.}

In Equation~\eqref{e:shell}, we neglect the reaction force on the
shell exerted by evaporation flows away from the ionization front
because it is not significant for embedded \HII regions.  The thrust
term would be important in driving expansion of classical,
blister-type \HII regions for which ionization fronts are kept
D-critical (\citealt{mat02, kru06}; \citetalias{KM09}).  We also
ignore the slowdown of the shell caused by turbulent ram pressure of
the external, neutral medium (e.g., \citealt{tre14,gee15}), which is
difficult to model within our spherically-symmetric, one-dimensional
model.

\begin{figure}
  \epsscale{1.2}\plotone{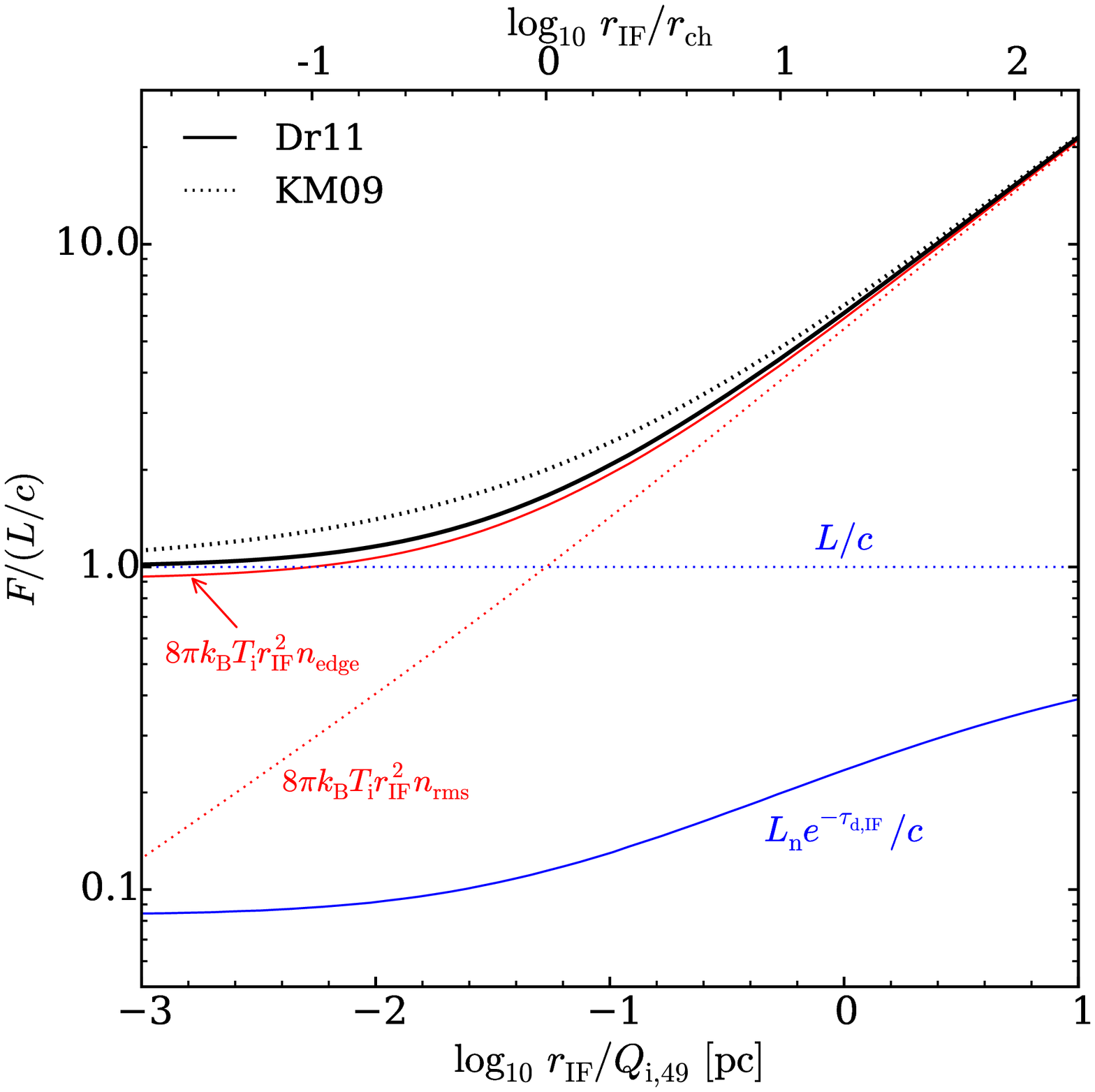} \caption{Outward forces on the shell
    surrounding the \HII region, shown as a function of the shell
    radius $\rsh = \rIF$, scaled as $\rIF/\Qiunit$ (lower $x$-axis) or
    $\rIF/\rch$ (upper $x$-axis). The solid lines denote the forces
    based on the quasi-static equilibrium model of \citetalias{Dr11},
    while the dotted lines are the effective forces assuming no
    attenuation of radiation inside the shell
    \citepalias[e.g.,][]{KM09}. The total forces and the radiative and
    thermal contributions are drawn as black, blue, and red curves,
    respectively. An \HII region with given $\Qi$ evolves from left to
    right. When $\rIF \ll \rch$, the radiation force pushes the gas
    within the \HII region outward, effectively communicating most of
    the radiation force $L/c$ to the surrounding shell through a
    dense, ionized shell ($\nedge/\nrms \gg 1$). As the \HII region
    expands to $\rsh > \rch$, direct radiation forces become
    unimportant and $\nedge \approx \nrms$.}\label{f:fout}
\end{figure}

Figure~\ref{f:fout} plots $\Fout$ as well as the respective
contributions of radiative and thermal pressures against the shell
radius $\rsh$, taken to equal $\rIF$, as the black, blue, and red
solid lines, respectively, utilizing the \citetalias{Dr11} solutions
with $\beta=1.5$ and $\gamma=11.1$. Apparently, the thermal term
dominates the radiative term by non-ionizing photons over the whole
range shown in Figure~\ref{f:fout}. This should {\it not} be
interpreted as an indication that radiation is unimportant for the
shell expansion. Rather, the effect of the radiation force is
indirectly communicated to the shell by increasing the ionized gas
density $\nedge$ above $\nrms$, with the limiting ratio $\nedge/\nrms$
given by the factor in Equation~\eqref{e:nedgelim} when
$\Qi\nrms\rightarrow \infty$.

The behavior of the outward forces for varying $\rIF$ can be readily
understood in comparison with the usual approximation adopted by
previous authors (e.g., \citealt{har09};
\citetalias{KM09,MQT10}). Assuming that all the photons from the
source are absorbed by the shell, that the \HII region has uniform
interior density $\nrms$, and that infrared photons re-radiated by
dust freely escape the system, the effective outward forces on the
shell from direct radiation and gas pressure are taken as
\begin{equation}\label{e:frad_eff}
 F_{\rm rad,eff}=L/c,
\end{equation}
and
\begin{equation}\label{e:fthm_eff}
 F_{\rm thm,eff}=8\pi\kB\Tion\nrms\rsh^2\,,
\end{equation}
respectively, where $\nrms$ is given in terms of $\Qi$, $\fion$, and
$\rsh=\rIF$ using Equation~\eqref{e:rIF}.  These together with the
total effective force $F_{\rm out,eff}=F_{\rm rad,eff}+F_{\rm
  thm,eff}$ are plotted as dotted lines in
Figure~\ref{f:fout}. Evidently, $F_{\rm out,eff}$ agrees well (within
20\%) with $\Fout$ from Equation \eqref{e:fout}.  Comparison of the
thermal term in Equation \eqref{e:fout} with Equation
\eqref{e:fthm_eff} shows that the enhanced thermal force for small
radius is due to increased $\nedge/\nrms$, which is caused by
radiation pressure in the \citetalias{Dr11} solutions.  At
sufficiently small radius, the total outward force is in fact equal to
$F_{\rm rad,eff}$.\footnote{Combining Equations~\eqref{e:rIF},
  \eqref{e:nedgelim}, and \eqref{e:fout}, one can show that $\Fout
  \rightarrow L/c$ in the limit of $\Qiunit\nrms \rightarrow \infty$.}

Since $F_{\rm thm,eff} \propto \nrms\rsh^{2} \propto
\fion^{1/2}\rsh^{1/2}$, while $F_{\rm rad,eff}$ remains constant, one
can write
\begin{equation}\label{e:fout_eff}
  F_{\rm out,eff} = \dfrac{L}{c} \left[ 1 +
    \left(\frac{\fion}{\fionch}\frac{\rsh}{\rch}\right)^{1/2} \right]\,,
\end{equation}
with the characteristic radius $\rch$ at which $F_{\rm rad,eff} =
F_{\rm therm,eff}$ defined by \citepalias{KM09}
\begin{equation}\label{e:rch}
\begin{split}
  \rch & \equiv \left(\dfrac{(1 + \beta)\sigmad}{\gamma}\right)^2
  \dfrac{\Qi}{12\pi\alphaB\fionch} \\
  & \rightarrow 5.3\times 10^{-2}\pc \;\Qiunit \,,
\end{split}
\end{equation}
where $\fionch\equiv\fion(\rch)$ is $0.32$ for our fiducial parameters
$\beta=1.5$ and $\gamma=11.1$.\footnote{\citetalias{KM09} adopted a
  constant value of $0.73$ for both $\fion$ and $\fionch$ for their
  approximate treatment.}  Expansion is driven predominantly by
radiation in the regime with $\rsh/\rch<1$, while ionized-gas pressure
is more important for $\rsh/\rch>1$. Thus, the relative importance of
radiation pressure within an \HII region can be assessed by the ratio
\begin{equation}\label{e:rIFrch}
\begin{split}
\dfrac{\rIF}{\rch} & = \fionch\left[
  \frac{\gamma(36\pi\alphaB)^{1/3}}{(1 + \beta)\sigmad}
  \right]^2\left(\dfrac{\fion}{\Qi^2 \nrms^2}
\right)^{1/3}\,. \\ &\rightarrow
\left(\dfrac{\fion}{0.32}\right)^{1/3}\left(\dfrac{\Qiunit\nrms}{2.55\times
  10^4 \cm^{-3}}\right)^{-2/3}\,,
\end{split}
\end{equation}
shown as a black dotted line in Figure~\ref{f:dr11_param} as a
function of $\Qiunit\nrms$.  Thus, shell expansion is dominated by
radiation pressure when $\Qiunit\nrms \gtrsim 10^4 \,\cm^{-3}$.

\subsection{Non-gravitating Similarity solutions}\label{s:asy}

In the absence of the inward gravitational forces ($\Fin=0$) and in
the limit of negligible core radius, it is straightforward to show
that Equations~\eqref{e:shell} and \eqref{e:fout_eff} yield the
similarity solutions
\begin{equation}\label{e:simil1}
  \nmean \rsh^4 =
  \dfrac{4-\krho}{2}\dfrac{L}{c}\dfrac{3}{4\pi\muH}t^2\,,
\end{equation}
in the limit of $\rsh/\rch \ll 1$, and
\begin{equation}\label{e:simil2}
  \nmean\rsh^{7/2} = \dfrac{3}{2}\dfrac{\kB\Tion}{\muH}
  \left(\dfrac{3\fion\Qi}{4\pi\alphaB}\right)^{1/2} \dfrac{(7 -
    2\krho)^2}{9-2\krho}t^2\,,
\end{equation}
in the limit of $\rsh/\rch \gg 1$, where $\fion$ is regarded as a
constant and $\muH = 2.34 \times 10^{-24} \gram$ is the mean atomic
mass per hydrogen and $\nmean = 3 \nc(\rsh/\rc)^{-\krho} /(3-\krho)$
is the mean number density interior to $\rsh$ (e.g., \citealt{kru06};
\citetalias{KM09}). Therefore, the shell radius and velocity vary with
time as $\rsh \propto t^{2/(4 - \krho)}$ and $\vsh \propto t^{(\krho -
  2)/(4 - \krho)}$ in the radiation-pressure driven limit, and $\rsh
\propto t^{4/(7 - 2\krho)}$ and $\vsh \propto t^{(2\krho -
  3)/(7-2\krho)}$ in the gas-pressure driven limit. \citetalias{KM09}
presented an analytic approximation valid in both limits by combining
Equations~\eqref{e:simil1} and \eqref{e:simil2}. Note that for
$\krho=3/2$, the velocity approaches a constant at sufficiently late
time.

\subsection{Valid Range of $\krho$}

When a cloud has too steep a density gradient, the shell mass becomes
smaller than the ionized gas mass within the ionization front. In this
case, the thin shell approximation we adopt would no longer be
valid. For instance, \citet{fra90} found that for a thermally-driven
\HII region formed in a cloud with $\krho=3/2$, the shock front moves
twice as fast as than the ionized sound speed, without significant mass
accumulation in the shocked shell. For $3/2 < \krho < 3$, an \HII
region becomes ``density bounded'' and develops ``champagne'' flows
rather than forming a shell (see also self-similar solutions by
\citealt{shu02}). To be consistent with our thin-shell approximation,
therefore, we consider clouds only with $\krho < 1.5$ in the following
analysis for shell expansion.

\begin{figure*}
\epsscale{1.}\plotone{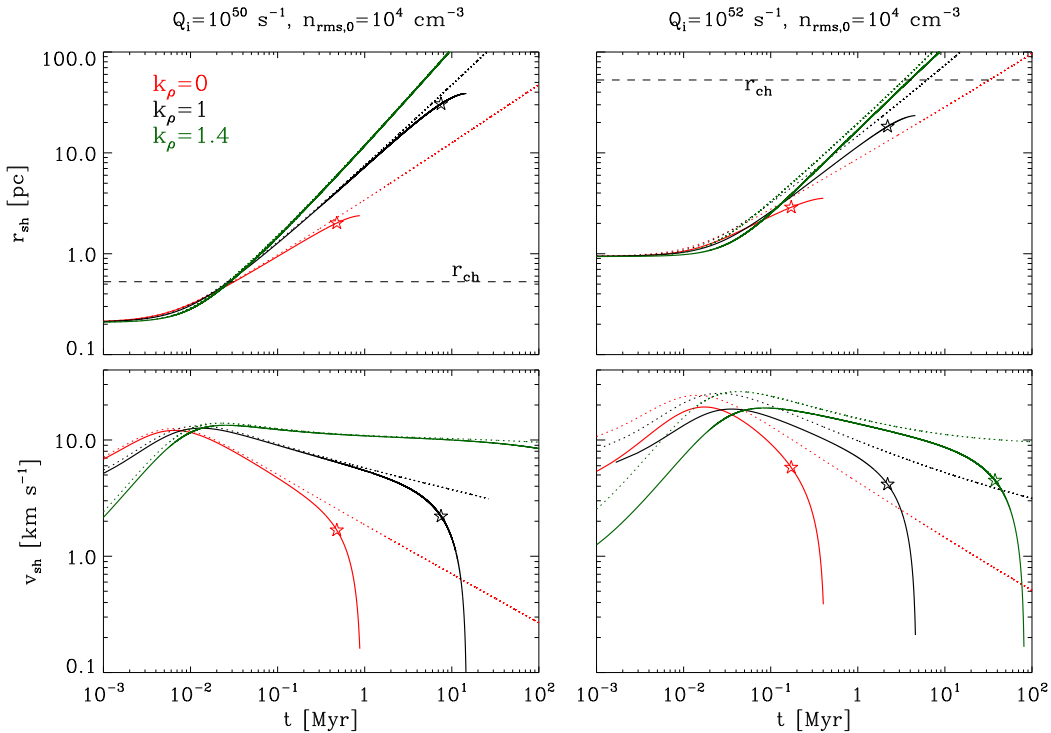}
\caption{Temporal evolution of the shell radius (upper) and the shell
  velocity (lower) for $\Qi = 10^{50} \second^{-1}$ (left) and $\Qi =
  10^{52} \second^{-1}$ (right). The initial rms density in the
  ionized region is fixed to $\nrmso=10^{4} \cm^{-3}$. The cases with
  $\krho=0$, 1, and 1.4 are plotted in red, black, and green. The
  solid and dotted lines correspond to the models with and without
  gravity, respectively.  The thin horizontal lines in the upper
  panels mark the characteristic radius $\rch$
  (Equation~\eqref{e:rch}), at which the dominant driving force
  switches from radiation to ionized-gas pressure. The star symbols
  indicate the radii where $\Ftot = 0$.}\label{f:ode}
\end{figure*}

\subsection{Shell Expansion}

When stellar gravity is included, we can relate $\Mstar$ to $\Qi$
through $\Mstar = \Qi/\Qitom$, where $\Qitom$ is the conversion factor
representing the ionizing photon output per unit stellar mass. The
corresponding light-to-mass ratio is $\Ltom = L/\Mstar= (1 + \beta)
\hnui \Qitom$. Obviously, these quantities depend on $\Mstar$ and vary
from cluster to cluster, especially for low-mass ones due to
stochastic fluctuations in the stellar population. In Appendix
\ref{s:ltom}, we perform Monte-Carlo simulations for the spectra of
coeval populations using the SLUG code \citep{kru15} to find spectral
properties as functions of $\Mstar$. Equations \eqref{e:ltom} and
\eqref{e:qitom} provide the fits to the resulting median values of
$\Ltom$ and $\Qitom$. While $\Ltom$ and $\Qitom$ saturate to constant
values for $\Mstar\gtrsim10^4\Msun$, they decrease sharply as $\Mstar$
decreases below $\sim 10^3\Msun$, due to a rapid decrease in the
number of O-type stars in the realizations of low-mass clusters. In
this work, we use Equation~\eqref{e:ltom} for conversion between
$\Mstar$ and $\Qi$, while fixing to $\beta=1.5$ and $\hnui=18\eV$: we
have checked that varying $\beta$ and $\hnui$ does not affect our
results much.

To integrate Equation~\eqref{e:shell}, we first choose a set of values
for $\beta$, $\gamma$, $\Qi$, $\nrmso$, and $\krho$. We then specify
$\fion$ from the \citetalias{Dr11} solutions and $\rIFo$ from
Equation~\eqref{e:rIF}. At $t=0$, a shell with a zero velocity is
located at $\rsh(0)=\rIFo$. The outside density profile is taken as
$n(r) = \nrmso(r/\rIFo)^{-\krho}$ for $r\geq \rsh$ and the shell mass
at any radius is given by $M_{\rm sh} =
4\pi\muH\nrmso\rsh^{3-\krho}\rIFo^{\krho} / (3-\krho)$.

As illustrative examples, we fix $\beta=1.5$, $\gamma=11.1$, and
$\nrmso = 10^4 \cm^{-3}$, and vary $\krho$ and $\Qi$.
Figure~\ref{f:ode} plots as solid lines the temporal behavior of the
shell radius (upper panels) and velocity (lower panels) for $\Qiunit =
10$ (left) and $\Qiunit = 10^3$ (right). The solutions without gravity
are compared as dotted lines. The cases with $\krho=0$, 1, and 1.4,
plotted in red, black, and green, respectively, show that the shell
expansion is faster in an environment with a steeper density profile,
because the shell mass increases more slowly for larger $\krho$.  The
horizontal dashed lines mark the the characteristic radii, $\rch=0.53$
and $53 \pc$ for $\Qiunit = 10$ and $10^3$, respectively, inside
(outside) of which shell expansion is driven primarily by radiation
(gas) pressure. The star symbols mark the radius where $\Ftot=0$,
beyond which expansion is slowed down by the shell self-gravity
significantly.

The solutions converge to the asymptotic power-law solutions discussed
in Section \ref{s:asy} (Equations~\eqref{e:simil1} and
\eqref{e:simil2}), although strong stellar gravity in the case with
$\Qiunit=10^3$ makes the shell expansion deviate from the
non-gravitating solutions early time.  Self-gravity of the swept-up
shell becomes important in the late stage, eventually halting the
expansion. The maximum shell velocity is only mildly supersonic with
respect to the ionized gas owing to the rapid increase in the shell
mass. Note that when $\Qiunit=10$, the driving force changes from
radiation to gas pressure at small $\rsh$, while shell expansion is
always dominated by radiation pressure when $\Qiunit=10^3$.

\subsection{Comparison with Numerical Simulations}\label{s:sim}

So far we have used a very simple model to study dynamical expansion of
a spherical shell surrounding an \HII region. In order to check how
reliable our results are, we run direct numerical simulations using the
\textit{Athena} code in spherical geometry \citep{sto08}, as described
in Appendix \ref{s:num}.  As an initial state, we consider a source of
radiation at the center of a radially stratified cloud with $\krho=1$.
To handle the transfer of radiation from the source, we adopt a
ray-tracing technique explained in \citet{mel06b} and \citet{kru07a}.
While gas inside the \HII region is evolved self-consistently, we
ensure that the outer envelope unaffected by radiation maintains its
initial hydrostatic equilibrium.

\begin{figure}
\epsscale{1.}\plotone{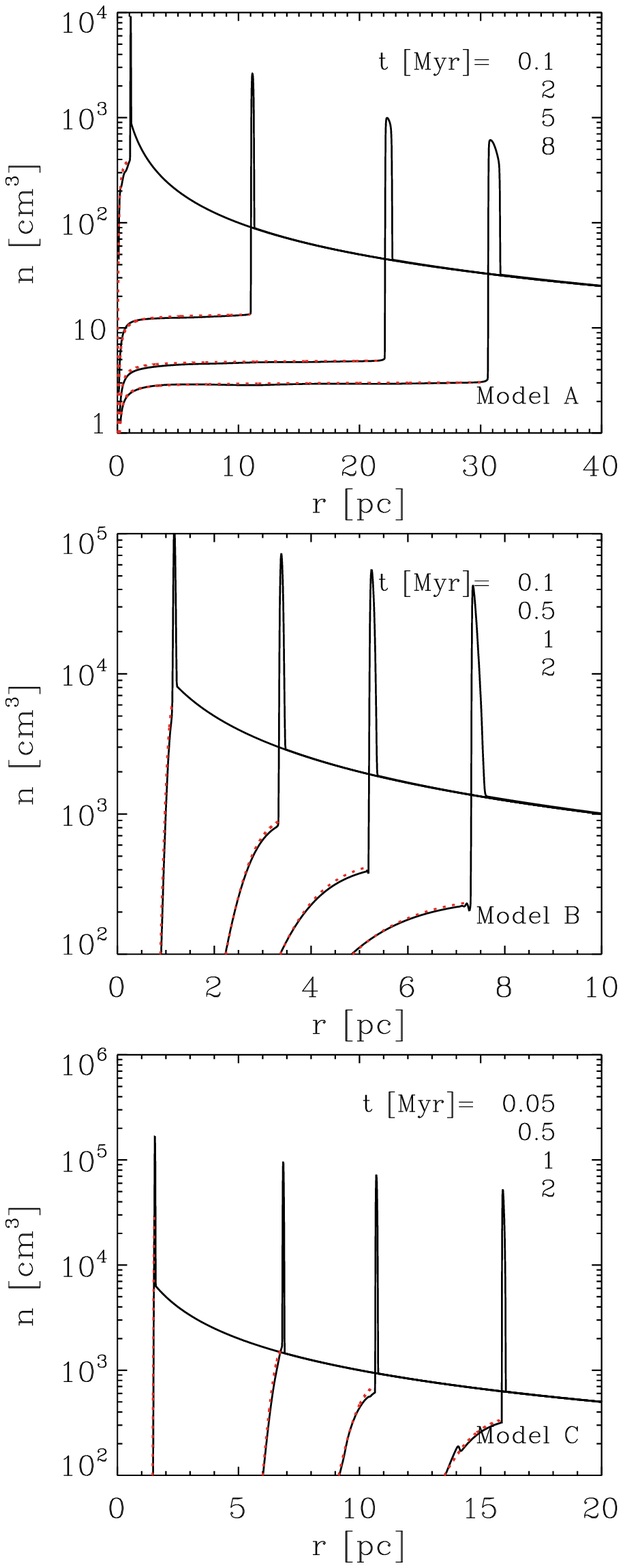} \caption{Evolution of the radial
  density distributions in numerical simulations of expanding \HII
  regions in a power-law density background with $\krho=1$ and $r_0 =
  1\pc$ for (a) Model A with $\Qi=10^{49} \second^{-1}$ and
  $n(r_0)=10^3 \cm^{-3}$, (b) Model B with $\Qi=10^{51} \second^{-1}$
  and $n(r_0)=10^4 \cm^{-3}$, and (c) Model C with $\Qi=10^{52}
  \second^{-1}$ and $n(r_0)=10^4 \cm^{-3}$. The red dotted lines plot
  the corresponding static equilibrium solutions of \citetalias{Dr11},
  describing the interior of the \HII region.}\label{f:sim1}
\end{figure}

Figure~\ref{f:sim1} plots as solid lines the radial density
distributions at a few epochs for Model A (with $\Qiunit=1$ and
$n(r_0)=10^3\cm^{-3}$; top), Model B (with $\Qiunit=10^2$ and
$n(r_0)=10^4\cm^{-3}$; middle), and Model C (with $\Qiunit=10^3$ and
$n(r_0)=10^4\cm^{-3}$; bottom), all with $\krho =1$ and $r_0=1 \pc$.
In Model A, the effect of radiation on shell expansion is almost
negligible since $\Qiunit\nrms < 10^4\cm^{-3}$. Its overall expansion
is in good agreement with the classical picture of thermally driven
expansion \citep[e.g.,][]{spi78}.  As soon as the ionization front
reaches the initial Str\"omgren radius ($\sim 0.3 \pc$), an isothermal
shock wave forms and propagates outward. At the same time, rarefaction
waves excited at the ionization front propagate radially inward
\citep[e.g.,][]{art11}, gradually turning into acoustic disturbances
that travel back and forth between the ionization front and the
center.  In this model, the density profile in the ionized region is
nearly flat.

On the other hand, Model C is dominated by radiation. Gas at small $r$
is pushed out supersonically to create a central cavity together with
an ionized shell within $t\sim 10^3 \yr$.\footnote{Assuming that the
  dust absorption dominates over the photoionization, the acceleration
  on dusty gas at $r$ is given by $L(r)\sigmad/(4\pi r^2c\muH)$. Then,
  the timescale for a gas parcel to travel over a distance $d$ in the
  ionized region is roughly $t\sim 10^3 (r^2 d/\rIFo^3)^{1/2} t_{\rm
    rec}$, where $t_{\rm rec} = (\alphaB n)^{-1}$ is the recombination
  timescale \citep{art04}.}  After the shell formation, the gas in the
ionized region relaxes into a quasi-static equilibrium state within a
sound crossing time across the \HII region. In Model B, the shell
expands mostly due to radiation pressure until it reaches $\rsh=6.3
\pc$ at $t=0.82\Myr$, after which the driving force switches to
ionized-gas pressure.

Figure~\ref{f:sim1} also plots as red dashed lines the corresponding
\citetalias{Dr11} quasi-static solutions inside the ionization front,
which are in good agreement with the results of the time-dependent
simulations. This not only confirms that the static solutions of
\citetalias{Dr11} are applicable even for expanding \HII regions, but
also validates our ray-tracing treatment of radiation in the numerical
simulations.

\begin{figure}
\epsscale{1.}\plotone{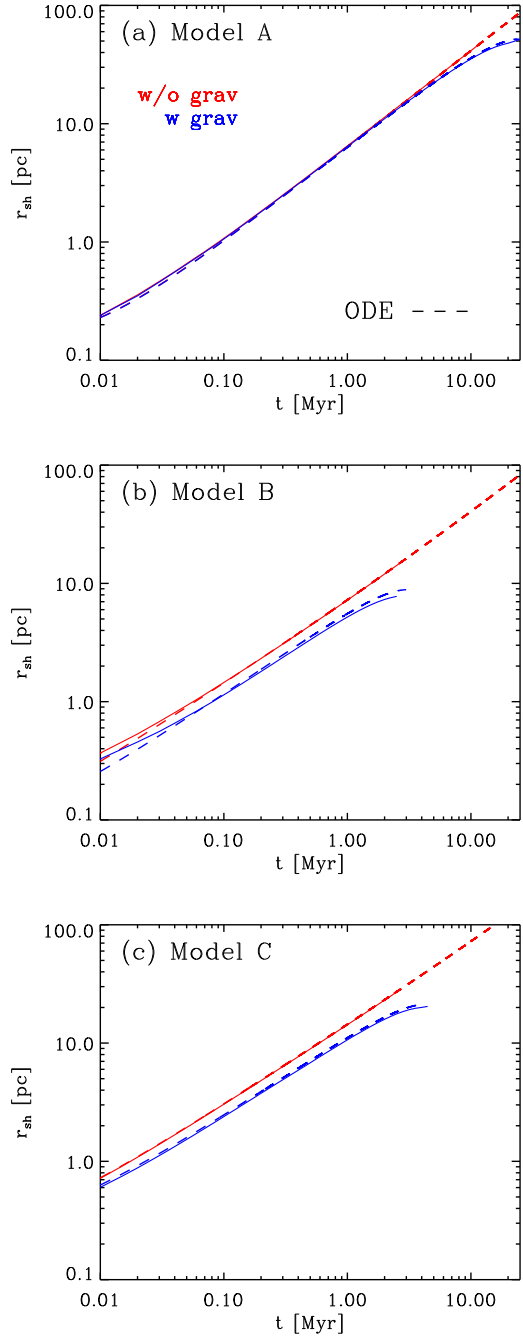} \caption{Temporal evolution of the
  shell radius in the simulations (solid lines) compared to the
  solutions of Equation~\eqref{e:shell} (dashed lines).  The cases
  with and without gravity are plotted in blue and red, respectively.
  Other than small differences at early time, the semi-analytic
  results are overall in good agreement with those of the
  simulations.}\label{f:sim2}
\end{figure}

Figure~\ref{f:sim2} plots the temporal changes (solid lines) of the
shell radius from the simulations in comparison with the solutions
(dashed lines) of Equation~\eqref{e:shell}. The cases with and without
gravity are plotted in blue and red, respectively.  Small differences
at early time between the results from the different approaches are
caused mainly by the assumption of a vanishing shell velocity at $t=0$
in solving Equation~\eqref{e:shell}, while the shell in simulations
has a non-zero velocity when it first forms.  Nevertheless,
semi-analytic and numerical solution agree with each other within
$\sim 5\%$ after $0.05 \Myr$, suggesting that our semi-analytic model
of shell expansion is quite reliable. Although the inclusion of
gravity does not make significant changes in the internal structure of
the ionized region, it is self-gravity that makes the swept-up shell
decelerate and eventually stall at $\rsh=51 \pc$ when $t = 23 \Myr$ in
Model A. Gravity becomes more significant for larger $\Qi\nrms$,
stopping the shell expansion at $t \approx 3$--$4 \Myr$ in Models B
and C, despite correspondingly stronger outward forces.

\section{Minimum Efficiency of Star Formation for Cloud
Disruption}\label{s:sfe}

Consider a star cluster comprising a mass fraction $\SFE$ of its birth
cloud, and located at the cloud center. The \HII region produced by
the luminous cluster launches an expanding shell that sweeps up the
bulk of the gas in the cloud as it grows. (see also, e.g.,
\citealt{har11}; \citetalias{MQT10}). If it is luminous enough, the
cluster is able to disrupt the entire parent cloud. For a low-mass
cluster, on the other hand, only a small volume near the center is
affected by the shell expansion, with the major portion of the cloud
remaining intact. Therefore, there exists a minimum star formation
efficiency $\SFEm$ such that a cloud with $\SFE < \SFEm$ is not
disrupted and will undergo further star formation, while a cloud with
$\SFE > \SFEm$ will be disrupted completely, with no further star
formation.  In real turbulent clouds, both star formation and gas mass
loss occur continuously (e.g., \citealt{dal12}, \citealt{ski15},
\citealt{ras15}), and one can expect $\SFE$ increases with time until
$\SFE\approx\SFEm$ at which point feedback is able to destroy or
evacuate all the remaining gas (see Section~\ref{s:dis} for more
discussion).

In this section, we use the simple model of shell expansion described
in Section \ref{s:exp} to evaluate $\SFEm$. \citet{fal10} studied how
gas removal due to various feedback processes regulates the efficiency
of star formation, and estimated $\SFEm$ for massive, compact
protoclusters by considering only the radiation pressure. Our model
extends their work by including both ionized gas pressure and gravity.
\citetalias{MQT10} integrated equations for shell expansion due to
radiation pressure, gas pressure, stellar winds, and protostellar
outflows, etc., as applied to several specific cases of massive
GMCs. Our aim here is to find the systematic dependence of $\SFEm$ on
cloud parameters such as mass, surface density, and density profile,
etc., which will be useful to assess the effectiveness of \HII region
feedback in a wider range of physical conditions.

\subsection{Fiducial Case}\label{s:fid}

Let us consider an isolated spherical cloud with mass $\Mcl$, radius
$\Rcl$, and mean surface density $\Sigmacl \equiv \Mcl/(\pi \Rcl^2)$. We
assume the cloud is gravitationally bound with the one-dimensional
turbulent velocity dispersion $\sigma = (\avir G\Mcl/5\Rcl)^{1/2}$,
where $\avir$ is the usual virial parameter of order unity
\citep[e.g.,][]{ber92}. At $t=0$, the cloud forms a star cluster with
mass $\Mstar = \SFE\Mcl$ instantaneously at its center. The remaining
gas mass $(1 - \SFE) \Mcl$ is distributed according to Equation
\eqref{e:iniden} with $r_{\rm c} = 0.1 \Rcl$ within the cloud radius
$\Rcl$; we have checked that the choice of the core radius has little
influence on the resulting $\SFEm$ as long as it is sufficiently
small. The cluster emits ionizing photons at a rate $\Qi =
\Qitom\Mstar$, producing a shell at $\rsh=\rIFo$, which we determine
by substituting $\nrms(<\rIFo)$ for $\nrms$ in Equation~\eqref{e:rIF}.

To determine $\SFEm$, we first take a trial value of $0 < \SFE < 1$
and integrate Equation~\eqref{e:shell} over time until the shell
reaches the cloud boundary. At $\rsh = \Rcl$, we check if the shell
meets one of the following four disruption criteria: (1) $\vsh(\Rcl) =
\vbind \equiv [(1 + \SFE)G\Mcl/\Rcl]^{1/2}$, corresponding to a
vanishing total (kinetic plus gravitational) energy of the shell
\citep[cf.][]{mat02, kru06}\footnote{As a disruption criterion,
  \citet{mat02} and \citet{kru06} took $\vsh(\Rcl)=v_{\rm esc} =
  \sqrt{2G\Mcl/\Rcl}$, which is the escape speed of an unbound test
  particle, rather than the minimum speed for unbinding a thin shell
  at $r=\Rcl$.}; (2) $\vsh(\Rcl)=\sigma$, corresponding to a stalling
of the shell expansion by turbulent pressure inside the cloud (e.g.,
\citealt{mat02}; \citetalias{KM09}; \citealt{fal10}); (3) $\vsh(\Rcl)$
equal to a large-scale turbulent ISM velocity dispersion, $\vturb \sim
7 \kms$ \citep{hei03}, corresponding to merging of the shell with the
diffuse ISM; and (4) $\Ftot(\Rcl)=0$, corresponding to a balance
between gravity and outward forces (e.g., \citealt{ost11};
\citetalias{MQT10}). If none of these conditions are satisfied or if
$\rsh$ is unable to expand to $\Rcl$, we return to the first step and
repeat the calculations by changing $\SFE$.

\begin{figure}
\epsscale{1.2}\plotone{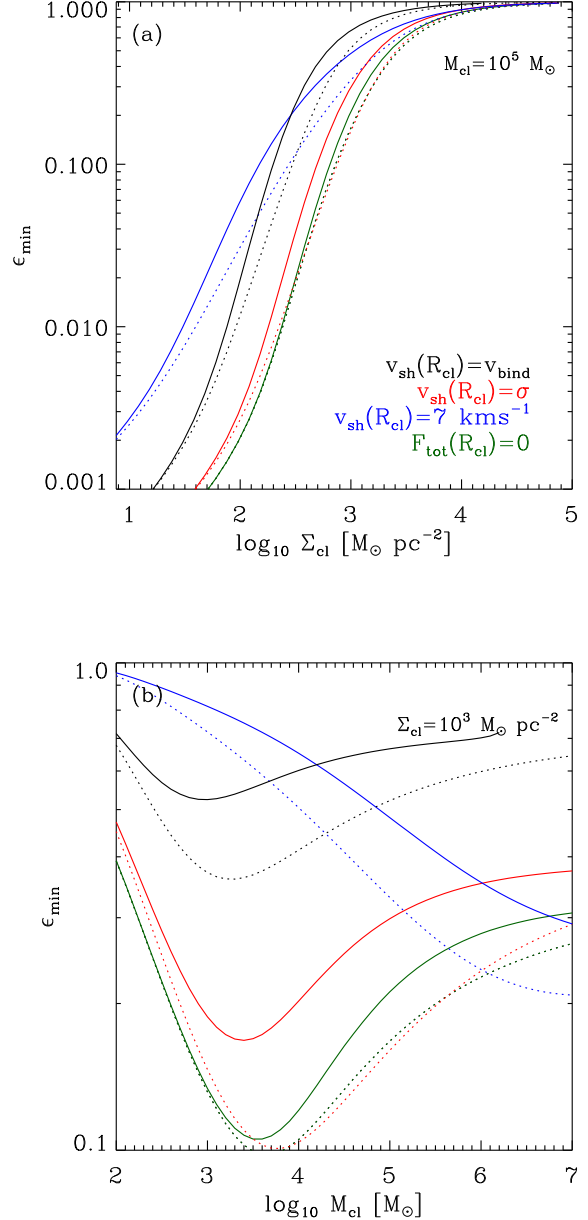} \caption{Minimum star formation
  efficiency $\SFEm$ for cloud disruption by \HII region expansion as
  a function of (a) the total mean surface density for fixed
  $\Mcl=10^5 \Msun$, and (b) the total mass for fixed $\Sigmacl = 10^3
  \Msun \pc^{-2}$, based on the four different criteria. Dotted lines
  show the analytic approximation using
  Equation~\eqref{e:vshsq}.}\label{f:SFE1}
\end{figure}

We take $\beta=1.5$, $\gamma=11.1$, $\avir = 1$, and $\krho=1$ for our
fiducial parameters.  Figure~\ref{f:SFE1} plots as solid lines $\SFEm$
(a) as a function of $\Sigmacl$ for clouds with $\Mcl = 10^5 \Msun$
and (b) as a function of $\Mcl$ for clouds with $\Sigmacl = 10^3 \Msun
\pc^{-2}$, for the four disruption criteria given above. As expected,
clouds with smaller surface density are more easily destroyed by
feedback, while more compact clouds need to have $\SFEm$ increasingly
closer to unity for disruption. The minimum efficiency depends more
sensitively on $\Sigmacl$ than $\Mcl$. In the case of the condition
$\vsh(\Rcl)=\vbind$, for instance, $\SFEm$ rises rapidly from $0.02$
to $0.66$ as $\Sigmacl$ increases from $10^{2}$ to $10^{3} \Msun
\pc^{-2}$ for fixed $\Mcl = 10^5 \Msun$, whereas it changes less than
a factor of two over $10^2 \Msun \leq \Mcl \leq 10^6 \Msun$ for fixed
$\Sigmacl = 10^3 \Msun \pc^{-2}$. Furthermore, a criterion based on a
higher shell velocity at the cloud boundary results in higher
$\SFEm$.

\begin{figure*}
\epsscale{1.0}\plotone{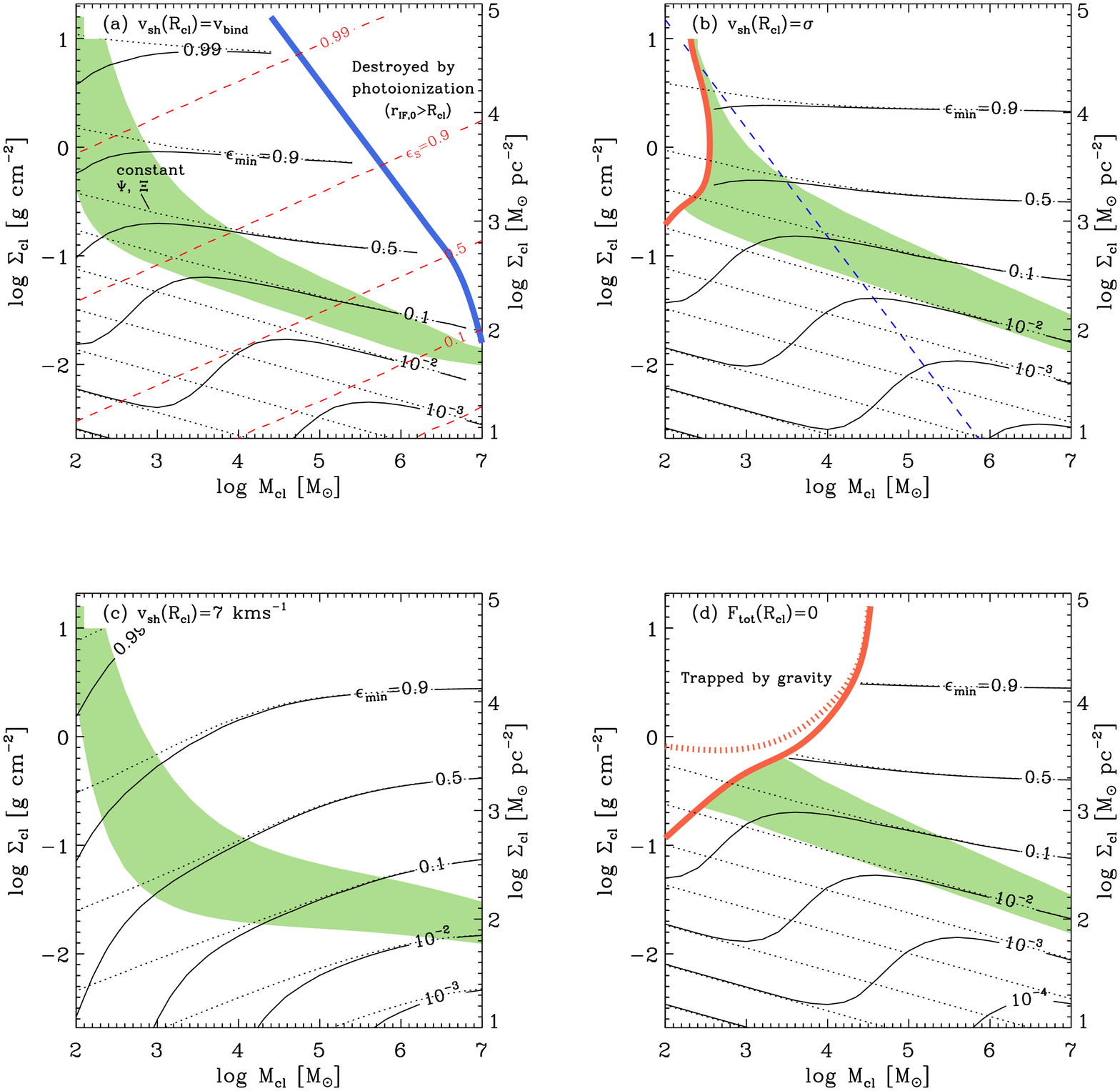} \caption{Contours of $\SFEm$ for
  cloud disruption by \HII region expansion, based on the criteria of
  (a) $\vsh(\Rcl)=\vbind=[(1+\SFE)GM/\Rcl]^{1/2}$, (b) $\vsh(\Rcl) =
  \sigma = (GM/5\Rcl)^{1/2}$, (c) $\vsh(\Rcl) = \vturb = 7 \kms$, and
  (d) $\Ftot (\Rcl) = 0$.  The solid and dotted contours correspond to
  variable and constant light-to-mass ratio, respectively. The red
  dashed contours in (a) plot the critical efficiency $\SFES$
  (Equation~\eqref{e:SFES}) for $\rIFo = \Rcl$, and the thick blue
  line draws the loci of $\SFEm = \SFES$, above which massive and
  compact clouds are subject to complete photodestruction. The upper
  left corner bounded by the thick red lines in (b) and (d)
  corresponds to ``trapped'' \HII regions for which a shell
  experiences infall rather than expansion. The shaded area with
  $\rIFo < \rch < \Rcl$ indicates the part of parameter space for
  which an \HII region undergoes a transition from radiation- to
  gas-pressure-driven expansion. Above the shaded region, $\Rcl <
  \rch$ so that the entire expansion is radiation-driven. Below the
  shaded region, $\rch<\rIFo$ and the entire expansion is driven by
  gas pressure. \citet{fal10} proposed a demarcation between
  radiation- and gas-pressure-dominated cases assuming $\SFE = 0.5$,
  shown as a blue dashed line in (b).}\label{f:SFE}
\end{figure*}

\begin{figure*}
\epsscale{1.0}\plotone{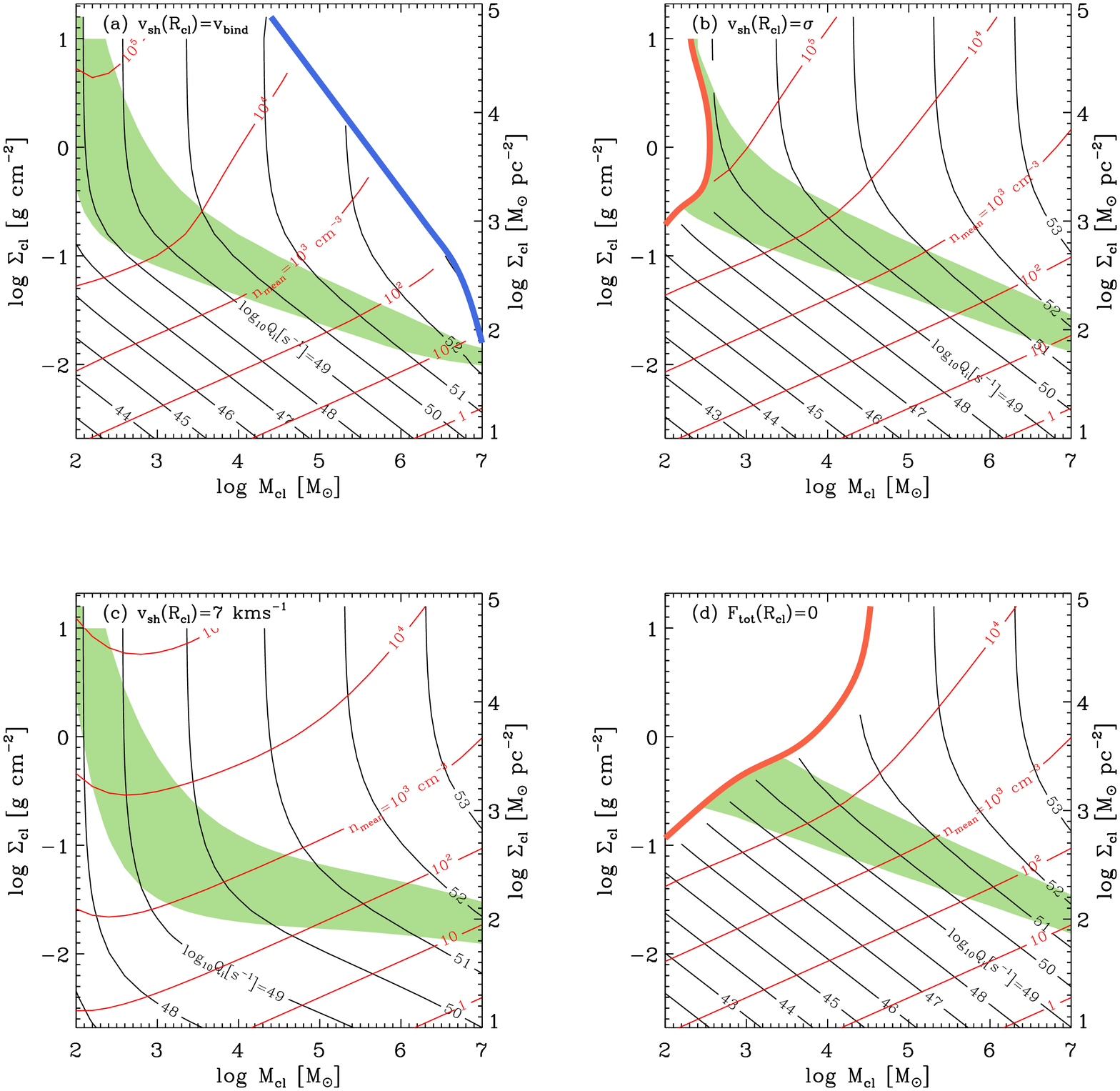}\caption{Contours of the rate of
  ionizing photons $\Qi$ (black) and the mean gas density $\nmean$
  (red) for clouds with $\SFE=\SFEm$.}\label{f:Qn}
\end{figure*}

Figure~\ref{f:SFE} plots the contours of $\SFEm$ on the
$\Mcl$--$\Sigmacl$ plane for the various criteria.  The solid curves
draw the results for the variable light-to-mass ratio (see
Equations~\eqref{e:ltom} and \eqref{e:qitom}), while the case of
constant $\Ltom= 943 \Lsun \Msun^{-1}$ and $\Qitom = 5.05 \times
10^{46} \second^{-1}\Msun^{-1}$ are shown as dotted curves. The
difference between the solid and dotted contours is caused by an
incomplete sampling of the initial mass function (IMF) which lowers
$\Ltom$ for low-mass clusters and thus yields higher $\SFEm$ for
$\Mstar \lesssim 10^3 \Msun$.  For the criterion $\vsh(\Rcl)=\vbind$,
typical GMCs with $10^4 \Msun \lesssim \Mcl \lesssim 10^6 \Msun$ and
$\Sigmacl \sim 10^2 \Msun \pc^{-2}$ in the Milky Way could be
destroyed by a handful of O stars ($\Qiunit \lesssim 10$) with
efficiency of $\lesssim 10\%$.  On the other hand, denser and more
compact cluster-forming clouds or clumps with $\Sigmacl \gtrsim 10^3
\Msun \pc^{-2}$ require the star formation efficiency larger than
$\sim 50\%$ for disruption.

In our model, the gas density is proportional to $(1 - \SFE)$ and the
ionizing rate $\Qi$ is proportional to $\SFE$, so that the size of the
\HII region becomes increasingly larger for higher $\SFE$. It is thus
possible to have $\rIFo \geq \Rcl$, indicating that the whole cloud is
completely photoionized from the beginning.  The upper right part
above the thick blue line in Figure~\ref{f:SFE}(a) corresponds to
these clouds destroyed by photoionization without involving shell
expansion. This can be seen more quantitatively by considering the
critical efficiency $\SFES$ that makes the initial Str\"{o}mgren
radius equal to $\Rcl$, for which Equation~\eqref{e:rIF} and
\begin{equation}
  \begin{split}
    \nrms(<\Rcl) &\approx (1 - \SFE) \left( 1 + \dfrac{\krho^2}{9 -
      6\krho} \right)^{1/2}
    \dfrac{3\pi^{1/2}}{4\muH}\dfrac{\Sigmacl^{3/2}}{\Mcl^{1/2}}\\ &\;\text{for}\;\;r_{\rm
      c} \ll \Rcl
  \end{split}
\end{equation}
yield
\begin{equation}\label{e:SFES}
  \SFES = \mathcal{C} - \sqrt{\mathcal{C}^2 - 1}\,,
\end{equation}
where
\begin{equation}
  \mathcal{C} = 1 + \dfrac{\fion\Qitom\muH^2}{2\pi^{1/2}\alphaB}
  \Mcl^{1/2}\Sigmacl^{-3/2}\,,\;\;\;\text{for}\;\;\krho=1\,.
\end{equation}
The red dashed contours in Figure~\ref{f:SFE}(a) plot $\SFES$,
demonstrating that the photodestruction boundary corresponds to the
loci of $\SFEm=\SFES$. For the other criteria, $\SFEm < \SFES$ for the
ranges of $\Mcl$ and $\Sigmacl$ shown in Figure~\ref{f:SFE}.

The region bounded by a thick red line (solid and dotted lines for
varying and fixed $\Ltom$ and $\Qitom$, respectively) in the upper
left corner of Figure~\ref{f:SFE}(b) and (d) corresponds to clouds
with $\Ftot(\rIFo) <0$ from the beginning or $\vsh = 0$ somewhere
before reaching the cloud boundary, indicating that a shell undergoes
infall rather than expansion due to strong gravity. In order for the
outward force to overcome gravity and drive expansion all the way to
the cloud surface, clouds in these regions must have star formation
efficiency higher than required by the respective criterion. To
describe these ``trapped'' \HII regions correctly, one needs to
consider accretion flows as well as gas rotation in a flattened
geometry, as in \citet{ket02,ket03,ket07}, which is beyond the scope
of the present paper.

Once we calculate the minimum efficiency required for cloud
disruption, we are positioned to derive various properties of \HII
regions and their host clouds.  Figure~\ref{f:Qn} plots contours of
$\Qi$ (black) as well as the mean gas number density in the cloud
(red) corresponding to $\SFEm$ for the various disruption
criteria. Clearly, larger $\Qi$ is necessary to destroy more massive
and compact clouds. The shaded regions shown in Figures~\ref{f:SFE}
and \ref{f:Qn} that run roughly diagonally from the upper left to
lower right corners correspond to \HII regions that make a transition
from the radiation-dominated to thermally-dominated regime in the
course of the expansion before disruption ($\rIFo < \rch <
\Rcl$).\footnote{The upper and lower boundaries of the shaded regions
  correspond to clouds with $\Rcl = \rch$ and $\rIFo = \rch$,
  respectively, for $\SFE=\SFEm$.}  For clouds with parameters above
(below) the shaded area, radiation (thermal) pressure plays the
dominant role throughout the expansion. As a demarcation line between
thermal and radiation pressure dominated \HII regions, \citet{fal10}
suggested $\Sigmacl/(1 \gram \cm^{-2}) = 0.15 \times [\Mcl/(10^4
  \Msun)]^{-1}$ by taking constant $\SFEm = 0.5$. This is plotted as a
blue dashed line in Figure~\ref{f:SFE}(b). As will be shown below, the
condition of $F_{\rm rad,eff}=F_{\rm thm, eff}$ at $\Rcl=\rch$ results
in $\Sigmacl \propto \SFE^{-2}\Mcl^{-1}$ in our model. Therefore,
since $\SFEm$ is a function of $\Sigmacl$ and $\Mcl$ instead of a
fixed value as assumed by \citet{fal10}, the shaded demarcation
differs from their proposal.

\begin{figure*}
\epsscale{1.0}\plotone{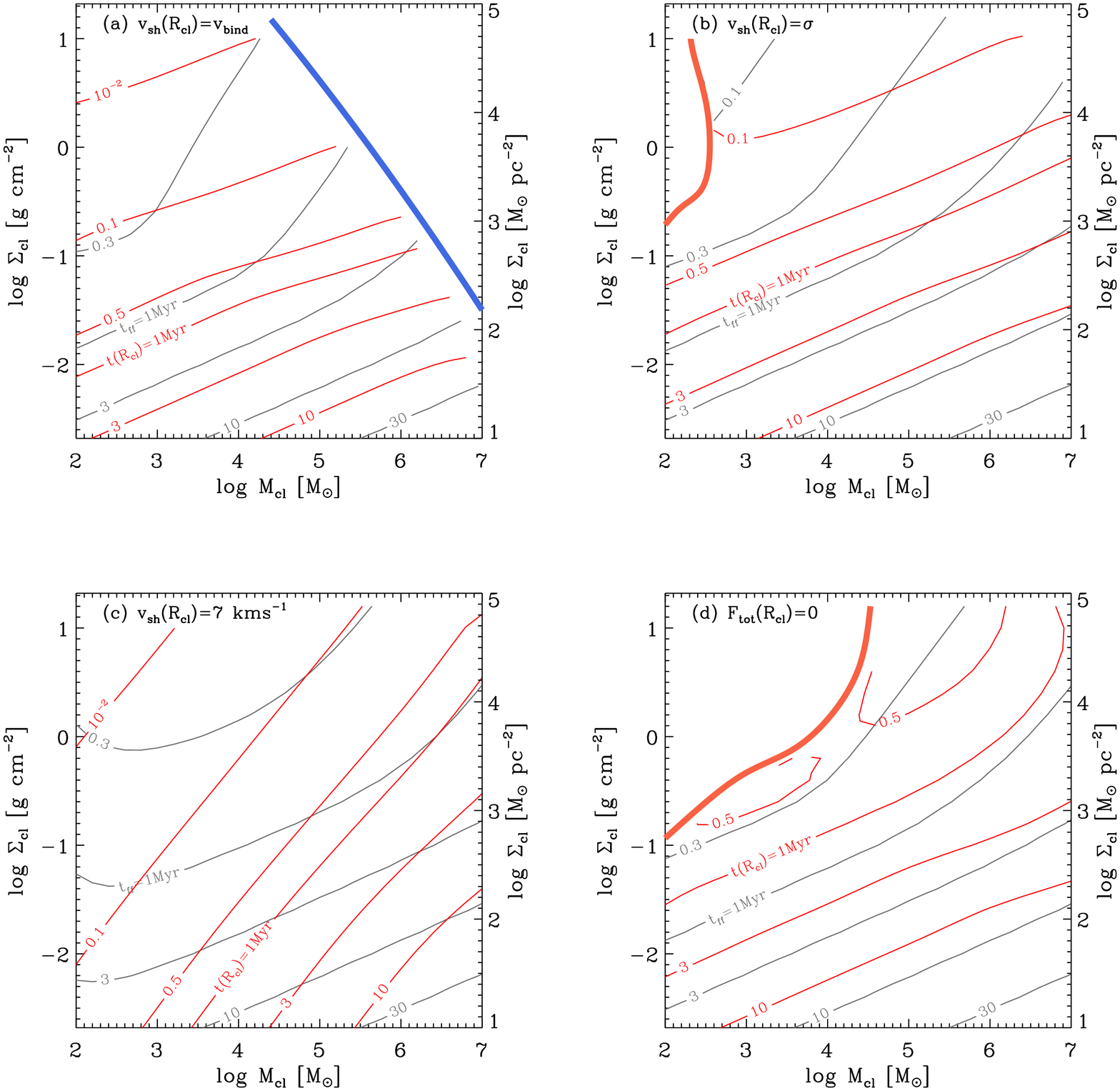}\caption{Contours of the expansion
  time $t(\Rcl)$ in red and the gas free-fall time $\tff$ in gray, for
  clouds with star formation efficiency $\SFE=\SFEm$.}\label{f:tR}
\end{figure*}

We also calculate the time $t(\Rcl)$ for the shell to move from $\rsh
= \rIFo$ to $\rsh=\Rcl$, plotted in Figure~\ref{f:tR} as red
contours. Note that $t(\Rcl)$ is comparable to or smaller than the gas
free-fall time $\tff =
(1-\SFEm)^{-1/2}(3\pi^{1/2}/8G)^{1/2}\Mcl^{1/4}\Sigmacl^{-3/4}$ shown
as gray contours. This is naturally expected for virialized clouds
with $\vsh(\Rcl) = \sigma$ or $\vsh(\Rcl) = \vbind$ since the shell
velocity approximately follows a power-law in time as $\vsh(\rsh)
\approx \vsh(\Rcl) (t/t(\Rcl))^{\alpha-1}$ before gravity takes over
(see Figure~\ref{f:ode}), with an exponent $2/3 < \alpha < 4/5$ for
$\krho=1$, yielding $t(\Rcl) = \alpha\Rcl/\vsh(\Rcl) \sim
\Rcl/\sigma$.  This suggests that cloud disruption by an expanding
\HII region is rapid, occurring roughly over the free-fall timescale,
provided the star formation efficiency is larger than the minimum
value. For the condition of $\vsh(\Rcl) = \vturb$, the slopes of the
$t(\Rcl)$-contours are different since $t(\Rcl) \propto \Rcl$.
Nevertheless, the shell expansion takes less than the free-fall time,
except for clouds with very large $\Sigmacl$ or very large $\Mcl$.

\subsection{Analytic Estimates}\label{s:ana}

One can deduce an approximate result for the minimum efficiency
analytically by utilizing the effective outward force given in
Equation~\eqref{e:fout_eff}.  Multiplying both sides of
Equation~\eqref{e:shell} by $\Msh\vsh$ and integrating the resulting
equation over time, we obtain
\begin{equation}\label{e:psh}
  \int d \psh^2 = 2 \int \Msh (F_{\rm out,eff}  - \Fin) d\rsh\,,
\end{equation}
where $\psh=\Msh\vsh$ is the shell momentum. Assuming $r_{\rm c} \ll
\Rcl$, $\fion=1$, and $\psh \rightarrow 0$ as $\rsh \rightarrow 0$,
Equation \eqref{e:psh} yields an expression for the shell velocity at
the cloud boundary as
\begin{equation}\label{e:vshsq}
\begin{split}
  \vsh^2(\Rcl) &= \eta_{\rm
    rad}\dfrac{\Ltom}{c}\dfrac{\SFE}{1-\SFE}\Rcl + \eta_{\rm thm}
      {\mathcal
        T}\dfrac{\SFE^{1/2}}{1-\SFE}\dfrac{\Rcl^{3/2}}{\Mcl^{1/2}}
      \\ & -
      \left[\eta_{\rm sh}(1-\SFE) + \eta_*\SFE\right]
      \dfrac{G\Mcl}{\Rcl}\,,
\end{split}
\end{equation}
where $\eta_{\rm rad}=2/(4-\krho)$, $\eta_{\rm thm}=4/(9-2\krho)$,
$\eta_{\rm sh}=1/(8-3\krho)$, $\eta_*=2/(5-2\krho)$, and ${\mathcal T}
= 8\pi\kB\Tion [(3\fion\Qitom)/(4\pi\alphaB)]^{1/2}$. Note that the
total force on the shell at $\rsh=\Rcl$ is approximately given by
\begin{equation}\label{e:f_Rcl}
  \Ftot \approx \dfrac{\SFE\Ltom\Mcl}{c} +
  \dfrac{{\mathcal T}\Mcl^{3/4}\SFE^{1/2}}{\pi^{1/4}\Sigmacl^{1/4}} - \dfrac{\pi
    G\Sigmacl \Mcl}{2}(1- \SFE^2)\,.
\end{equation}
The first and second terms in the right-hand-side of Equation
\eqref{e:f_Rcl} represent the radiative and thermal pressure forces,
respectively, while the last term comes from the total gravity. The
ratio of the first to second term is proportional to
$\sim(\SFE^2\Sigmacl\Mcl)^{1/4}$, suggesting that the relative role of
radiation pressure to thermal pressure depends not only on the cloud
mass and size but also on the star formation efficiency. The relation
in Equation~\eqref{e:vshsq} may be combined with the disruption
criterion $\vsh(\Rcl)=\vbind$, $\vsh(\Rcl)=\sigma$, and
$\vsh(\Rcl)=\vturb$ to obtain estimates for $\SFE$. Similarly,
Equation~\eqref{e:f_Rcl} may be used with the disruption criterion
$\Ftot(\Rcl)=0$. The estimates for the minimum efficiencies are
plotted as dotted lines in Figure~\ref{f:SFE1}. At high and low
$\Sigmacl$, the analytic estimates are very close to the solutions of
the ODE (integrating Equation~\eqref{e:shell}), although they can
depart by up to a factor of two between $\Sigmacl=10^2-10^3 \Msun
\pc^{-2}$.

\subsubsection{Radiation-pressure-driven Limit}

When the expansion is dominated by radiation pressure, we can keep
only the first term in the right-hand-side of Equation \eqref{e:vshsq}
to obtain
\begin{equation}\label{e:sferad1}
  \dfrac{\SFEm}{1 - \SFEm^2} = \dfrac{\pi
  Gc}{\eta_{\rm rad}\Ltom}\Sigmacl\,,\;\;\text{for}\;\; \vsh(\Rcl)=\vbind\,,
\end{equation}
and
\begin{equation}\label{e:sferad2}
  \dfrac{\SFEm}{1 - \SFEm} = \dfrac{\pi\avir
    Gc}{5\eta_{\rm rad}\Ltom}\Sigmacl\,,\;\;\text{for}\;\;
    \vsh(\Rcl)=\sigma\,.
\end{equation}
These clearly show that the minimum efficiency increases with the mean
surface density, independent of the mass. This is consistent with the
results of \citet{fal10}, although their effective outward force
includes the trapping factor accounting for hot stellar winds and
dust-reprocessed radiation. Similarly, a criterion based on
radiation-gravity force balance at the cloud boundary in Equation
\eqref{e:f_Rcl} yields
\begin{equation}\label{e:SFE_rad}
  \dfrac{\SFEm}{1 - \SFEm^2} = \dfrac{\Sigmacl}{2\Ltom/(\pi
    Gc)}\,,\;\;\text{for}\;\; \Ftot=0\,,
\end{equation}
independent of $\Mcl$ again (see also \citetalias{MQT10} and
\citealt{ras15}).

\subsubsection{Gas-pressure-driven Limit}

For expansions driven primarily by ionized-gas pressure, Equations
\eqref{e:vshsq} and \eqref{e:f_Rcl} give
\begin{equation}\label{e:sfet}
  \dfrac{\SFEm}{(1 - \SFEm^2)^2} = \left(\dfrac{\pi^{5/4}G}{\eta_{\rm
      thm}
    \mathcal{T}}\right)^2\Mcl^{1/2}\Sigmacl^{5/2}\,,\;\;\text{for}\;\;
  \vsh(\Rcl)=\vbind\,,
\end{equation}
\begin{equation}\label{e:sfetherm}
  \dfrac{\SFEm}{(1 - \SFEm)^2} = \left(\dfrac{\pi^{5/4}\avir
    G}{5\eta_{\rm thm}\mathcal{T}}\right)^2
  \Mcl^{1/2}\Sigmacl^{5/2}\,,\;\;\text{for}\;\; \vsh(\Rcl)=\sigma\,,
\end{equation}
and
\begin{equation}\label{e:SFE_therm}
  \dfrac{\SFEm}{(1-\SFEm^2)^2} = \left(\dfrac{\pi^{5/4}G}{2{\mathcal
      T}}\right)^2 \Mcl^{1/2}\Sigmacl^{5/2}\,,\;\;\text{for}\;\;
  \Ftot=0\,.
\end{equation}
Equations~\eqref{e:sfet}--\eqref{e:SFE_therm} imply $\SFEm \propto
\Mcl^{1/2}\Sigmacl^{5/2}$ for $\SFEm \ll 1$.

\subsection{Effects of $\krho$ and Trapped Infrared Radiation}\label{s:krhoIR}

In Section~\ref{s:fid}, we considered a stratified cloud with
$\krho=1$ and assumed that infrared photons emitted by dust freely
escape the cloud.  Here we relax these two constraints to calculate
$\SFEm$ in more general situations.

\begin{figure}
\epsscale{1.3}\plotone{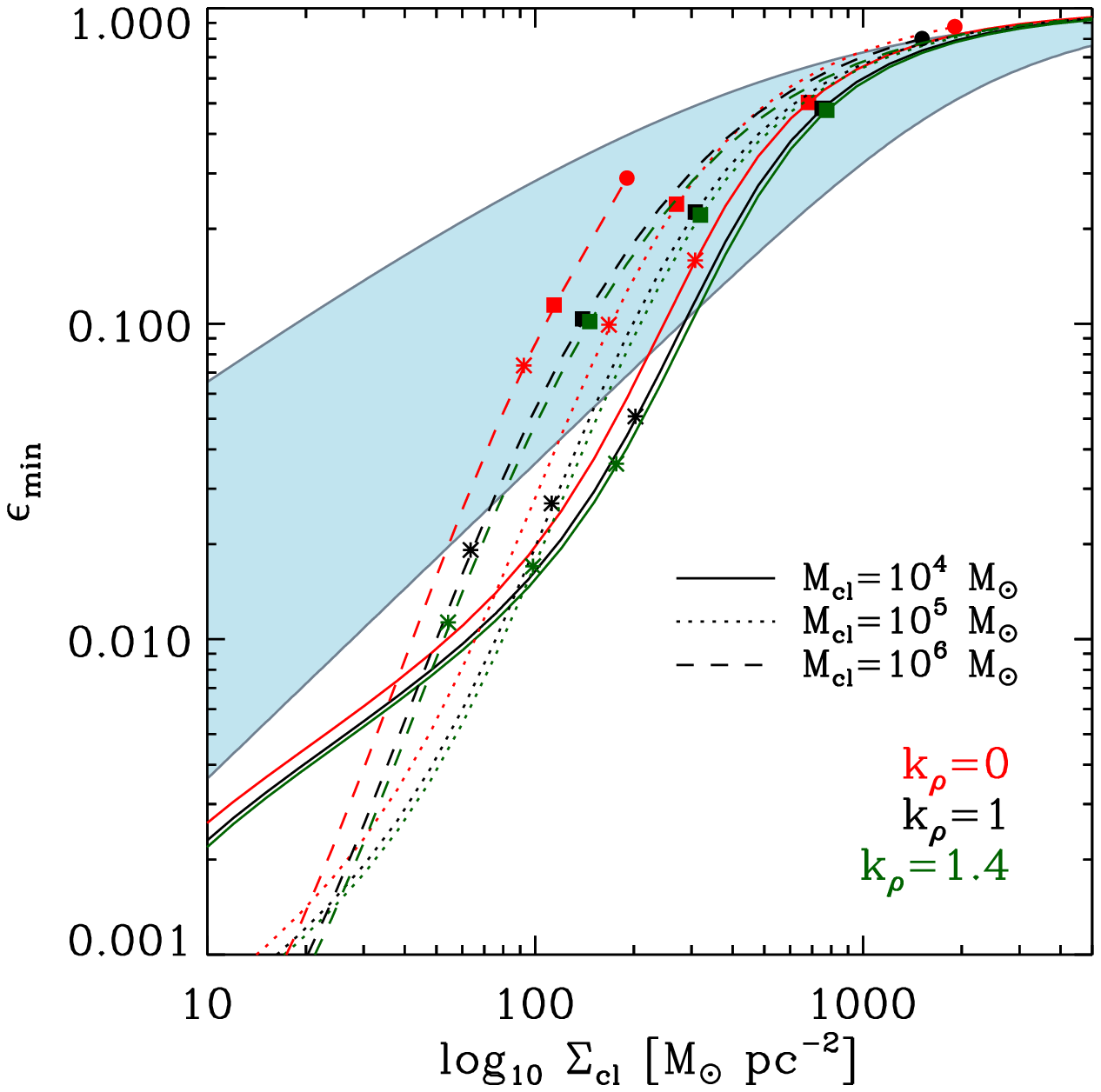}\caption{Effects of the power-law
  index $\krho$ on $\SFEm$ for the disruption criterion $\vsh(\Rcl) =
  \vbind$. Three cloud masses $\Mcl=10^4, 10^5, 10^6\Msun$ are
  chosen. Solid circles mark the positions where $\SFEm=\SFES$, beyond
  which entire clouds are photoionized even before shell
  expansion. Asterisks and solid squares indicate the positions where
  $\rch =\rIFo$ and $\rch = \Rcl$, respectively. Note that smaller
  $\krho$ increases $\SFEm$ but only slightly. The blue shaded area
  delineates the net star formation efficiency of turbulent clouds
  predicted by \citet{ras15}, which includes radiation pressure but
  ignores gas pressure, for a range of variance $0<\sigma_{\ln
    \Sigma}<1$ of the lognormal surface density distribution (see
  Discussion in the text).}\label{f:SFE_krho}
\end{figure}

\begin{figure*}
\epsscale{1.0}\plotone{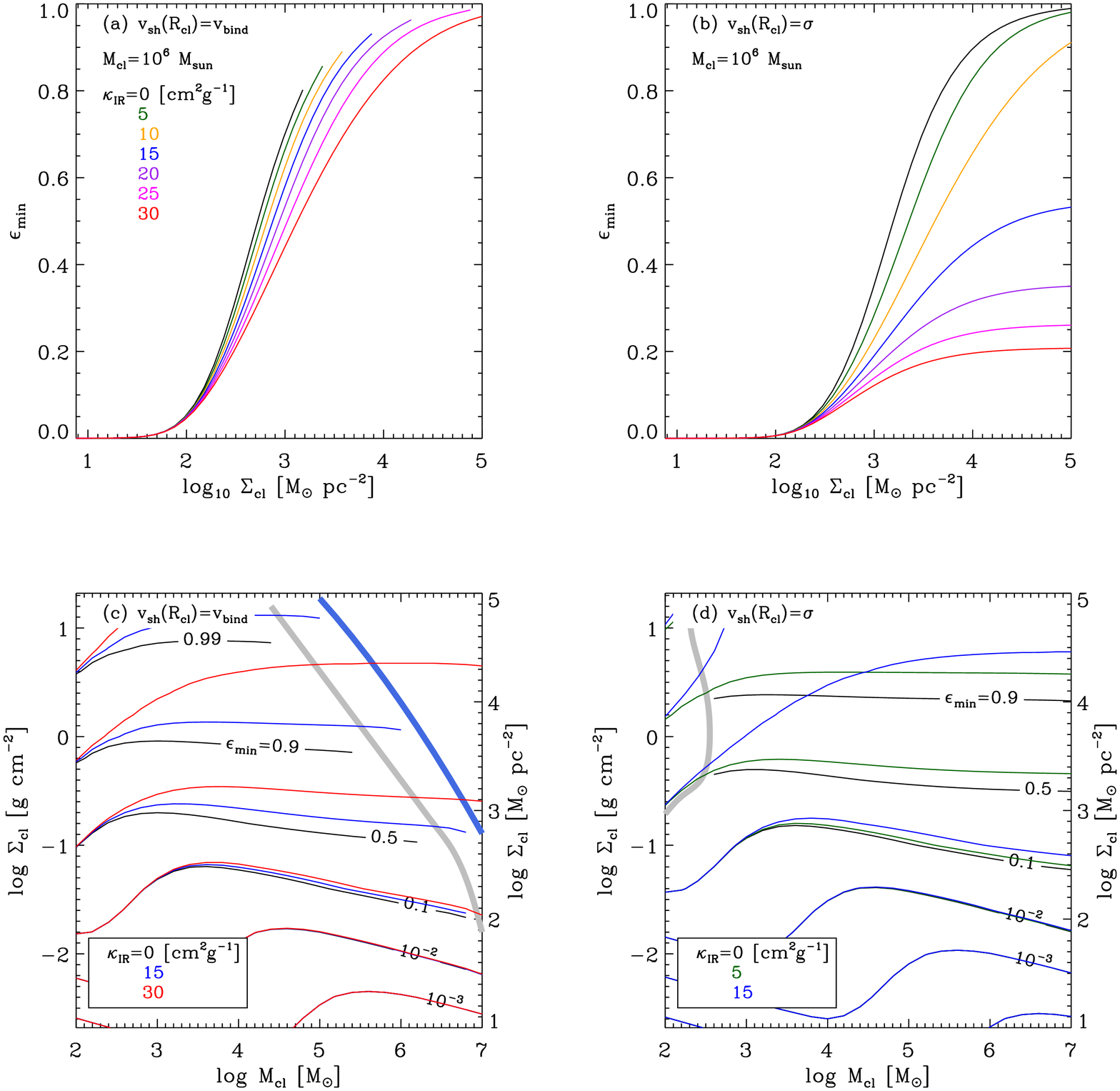} \caption{Effects of trapped infrared
  radiation on $\SFEm$ for the disruption criteria $\vsh(\Rcl)=\vbind$
  (left) and $\vsh(\Rcl)=\sigma$ (right). Upper panels: $\SFEm$ as a
  function of $\Sigmacl$ for $\kappaIR=0,5,\cdots,30 \cm^2
  \gram^{-1}$, when the cloud mass is $\Mcl=10^6\Msun$. Lower panels:
  Contours of $\SFEm$ in the $\Mcl$--$\Sigmacl$ plane. The thick lines
  in panel (c) draw the photodestruction boundaries for $\kappaIR=0$
  and $15\cm^2\gram^{-1}$, on which $\SFEm=\SFES$. The upper left
  region in panel (d) bounded by the thick line experiences infall
  rather than expansion.}\label{f:SFE_IR}
\end{figure*}

First, we explore the cases with differing $\krho$.  \HII regions in
clouds with a larger central density tend to be denser and smaller in
size initially. As a result, the radiation pressure becomes more
important in the initial expansion phase for larger $\krho$.
Figure~\ref{f:SFE_krho} plots $\SFEm$ as a function of $\Sigmacl$ for
$\krho=0$, 1, and 1.4 in the case of the disruption criterion
$\vsh(\Rcl)=\vbind$. Three cloud masses $\Mcl=10^4, 10^5, 10^6\Msun$
are chosen. Although clouds with steeper density profile have smaller
$\SFEm$ and are thus more readily disrupted, $\SFEm$ does not exhibit
strong dependence on $\krho$. In fact, Equation~\eqref{e:vshsq}
suggests that the shell momentum depends weakly on $\krho$ only
through the $\eta$ coefficients.

Next, we examine the effect of dust reprocessed infrared radiation. In
a high-column cloud optically thick to infrared radiation,
dust-reprocessed radiation can be trapped within the cloud, providing
significant boost to \HII region expansion (e.g., \citetalias{KM09,
  MQT10}; \citealt{hop11, ost11}; \citealt{ski15}). For the momentum
injection rate by the trapped radiation, we adopt the usual simple
prescription
\begin{equation}
  \FIR = \tauIR \frac{L}{c}\,,
\end{equation}
applicable to smooth and spherical clouds. Here, $\tauIR=\kappaIR
(\Sigma_{\rm sh} + \int_{\rsh}^{\Rcl} \rho(r) dr)$ is the infrared
optical depth through the cloud, with $\kappaIR$ being the Rosseland
mean dust opacity (treated as a constant for simplicity).\footnote{The
  reader is referred to \citet{ski13} for a more rigorous treatment of
  spherical shell expansion driven by trapped radiation (without
  gravity).}  Note that we include the contribution to $\tauIR$ from
the portion of the cloud outside the shell as well, corresponding to
the maximum possible efficiency of feedback by the trapped
radiation. As $\rsh$ increases, $\tauIR$ experiences a moderate
decrease (by a factor of three from $\rsh=0$ to $\rsh=\Rcl$ for
$\krho=0$).

We integrate Equation \eqref{e:shell} by adding $\FIR$ to $\Fout$.
Figure~\ref{f:SFE_IR} plots $\SFEm$, determined from two disruption
criteria, $\vsh(\Rcl)=\vbind$ (left panels) and $\vsh(\Rcl)=\sigma$
(right panels), for several different values of $\kappaIR$, as
functions of $\Sigmacl$ for $\Mcl=10^6\Msun$ (upper panels) and as
contours in the $\Mcl$--$\Sigmacl$ plane (lower panels). The other
parameters are the same as in the fiducial case.  Clearly, $\kappaIR$
tends to reduce $\SFEm$ since clouds are more easily disrupted. But,
its effect is significant only for sufficiently large $\kappaIR$
and/or sufficiently massive, high-column clouds so as to have $\rsh
\lesssim \rch$ and $\tauIR \gtrsim 1$. When $\kappaIR \approx 5 \cm^2
\gram^{-1}$ appropriate for solar-metallicity gas \citep{sem03},
$\SFEm$ is insensitive to $\FIR$. For larger values of $\kappaIR$, the
minimum efficiency is reduced considerably and the photodestruction
regime becomes less extended, as Figure~\ref{f:SFE_IR}(c) shows. For
clouds with $\tauIR \gg 1$, the trapped radiation dominates the direct
radiation ($\FIR > L/c$). In this case, $\SFEm$ computed from the
criterion $\vsh(\Rcl)=\sigma$ becomes close to $\varepsilon_{\rm
  min,IR} = [\Ltom\kappaIR/(2\pi cG) - 1]^{-1}$, a prediction from the
force balance between $\FIR$ and gravity \citep{ost11}. The
simulations of \citet{ski15} consider the evolution of turbulent
clouds in the limit $\tauIR > 1$ where reprocessed radiation dominates
direct radiation.

\section{Summary and Discussion}\label{s:sum}

\subsection{Summary}

Young massive stars have dramatic effects on the surrounding ISM
through ionizing radiation, winds, and supernova explosions. \HII
regions created by ionizing radiation from young star clusters
embedded in molecular clouds are able to destroy their natal clouds
under some circumstances, preventing further star formation in
them. In this paper we have used a simple semi-analytic model as well
as hydrodynamic simulations to study dynamical expansion of a dusty
\HII region around a star cluster and its role in cloud disruption.
Our expansion model is one-dimensional, assuming spherical symmetry,
and treats the structure of the ionized region using the solution of
\citetalias{Dr11}. We assume that as the \HII region expands, the
swept-up shell formed at the interface between the ionized region and
surrounding neutral neutral gas remains very thin. We solve the
shell's temporal evolution subject to outward contact forces arising
from radiation and thermal pressures, and inward gravity from the
cluster and the shell.

In our model, radiation pressure affects the shell expansion
indirectly through the enhanced thermal pressure at the ionization
front, a feature of the \citetalias{Dr11} solutions. The total outward
contact force in our detailed model agrees within $\sim20\%$ with the
combination of the effective radiation force $F_{\rm rad,eff}=L/c$ and
the effective gas pressure force $F_{\rm
  thm,eff}=8\pi\kB\Tion\rsh^2\nrms$ used in the simplified model
\citetalias{KM09} (see Figure~\ref{f:fout}).  Since $F_{\rm rad,eff}$
is constant while $F_{\rm thm, eff}$ depends on the shell radius
$\rsh$ as $F_{\rm thm, eff}\propto \rsh^{1/2}$ (apart from the weak
dependence on $\fion$), expansion is driven primarily by radiation
pressure when $\rsh < \rch$ (or equivalently when $\Qiunit\nrms
\gtrsim 10^4 \cm^{-3}$), and by thermal pressure when $\rsh >
\rch$. Here $\rch$ is the characteristic radius where $F_{\rm
  rad,eff}=F_{\rm thm, eff}$ (Equation~\eqref{e:rch}). We note that in
practice, radiation forces are conveyed to the surrounding neutral
shell indirectly, by compressing the ionized gas strongly to the outer
portion of the \HII region and increasing the density immediately
interior to the shell.  The cluster gravity is important in the early
phase of shell expansion, especially in the radiation-driven limit,
while shell self-gravity eventually halts expansion, after about the
free-fall timescale of the whole cloud.

To validate the assumptions used in our expansion model, we also
perform direct numerical simulations of expanding spherical \HII
regions for sample cases. We find that despite the presence of
small-amplitude acoustic disturbances, the radial density structure of
the ionized region in our numerical simulations is overall in good
agreement with the static equilibrium solutions of \citetalias{Dr11}
throughout the shell expansion. This justifies the use of the
\citetalias{Dr11} solutions even for expanding \HII regions. The
temporal changes in the shell position and velocity in our expansion
model also agree with the results of the numerical simulations until
gravity completely stops the expansion (Figures~\ref{f:sim1} and
\ref{f:sim2}).

Using our expansion model, we explore requirements for cloud
disruption by an expanding \HII region around a star cluster in
diverse star-forming environments, as characterized by the total cloud
mass $\Mcl$ and the mean surface density $\Sigmacl$. We also allow for
power-law internal density profiles in the cloud. As criteria of cloud
disruption, we consider the following four conditions: (1) the shell
velocity at the cloud boundary $\Rcl$ is reduced to
$\vsh(\Rcl)=\vbind= [(1+\SFE)G\Mcl/\Rcl]^{1/2}$, corresponding to a
vanishing total energy; (2) $\vsh (\Rcl)=\sigma$, the one-dimensional
internal velocity dispersion of the cloud; (3) $\vsh (\Rcl)=7\kms$,
the mean turbulent velocity dispersion of the diffuse ISM; and (4) the
net force on the shell at the cloud boundary is zero. As a function of
$\Mcl$ and $\Sigmacl$, we calculate the minimum efficiency of star
formation $\SFEm$ needed to satisfy each of the four conditions above
(Figures~\ref{f:SFE1} and \ref{f:SFE}). Using the effective outward
forces, we also derive analytic expressions for $\SFEm$ in the
radiation- and gas-pressure-driven limits (Section~\ref{s:ana}).

Based on the first criterion, $\vsh(\Rcl)=\vbind$ (see
Figure~\ref{f:SFE}a), GMCs in normal disk galaxies (typically $10^4
\Msun \lesssim \Mcl \lesssim 10^6 \Msun$ and $\Sigmacl \sim 50$--$200
\Msun \pc^{-2}$) can be destroyed with $\SFEm \lesssim 10 \%$. For
$\Mcl \le 10^5 \Msun$ and $\Sigma \le 10^2 \Msun \pc^{-2}$, expansion
is primarily due to gas pressure, whereas both gas and radiation
pressure are important at larger $\Mcl$ and $\Sigmacl$.
Disruption of cluster-forming clumps ($\Mcl
\gtrsim 10^3 \Msun$ and $\Sigmacl \gtrsim 10^3 \Msun \pc^{-2}$, from
Fig.~1 of \citealt{tan14}) requires a significantly higher efficiency
of $\SFEm \gtrsim 50 \%$, and is mainly driven by radiation
pressure. Massive clouds in starbursts ($\Mcl \gtrsim 10^5 \Msun$ and
$\Sigmacl \gtrsim 1 \gram \cm^{-2}$) would need to convert most of
their gas into stars ($\SFEm \gtrsim 0.9$) for the \HII region to
achieve disruption by direct radiation forces.
The minimum efficiency for the criterion
$\vsh(\Rcl) = \sigma$ is only slightly higher than that required from
$\Ftot(\Rcl)=0$, implying that the gravity is already taking over when
the shell velocity drops to $\sigma$, naturally expected for a
virialized cloud.

Star clusters with masses below $\sim 10^3 \Msun$ are likely to have
fewer OB stars and thus a smaller light-to-mass ratio than the
predictions of a fully sampled IMF (see Appendix~\ref{s:ltom}). The
required $\SFEm$ increases accordingly for these low-mass
systems. Since the size of \HII regions increases with increasing
efficiency, there exists a critical efficiency $\SFES$
(Equation~\eqref{e:SFES}) at which the initial Str\"{o}mgren radius is
equal to the cloud size. Clouds with $\SFEm \geq \SFES$ are regarded
as being disrupted by photoionization rather than shell
expansion. Such clouds are located in the upper right corner in the
$\Mcl$--$\Sigmacl$ plane in the case of the criterion $\vsh(\Rcl) =
\vbind$ (Figure~\ref{f:SFE}(a)). Under the criterion $\vsh(\Rcl) =
\sigma$ or $\Ftot(\Rcl)=0$, shells in the upper left corner are
subject to gravitational infall rather than expansion.  For clouds
disrupted by shell expansion, the disruption time is typically
comparable to or smaller than the gas free-fall time
(Figure~\ref{f:tR}), implying that the disruption is rapid once an
\HII region is formed.

We also examine the effect of differing cloud density profiles by
varying the power-law density index $\krho$. When all other quantities
are held fixed, momentum deposition is only slightly larger for clouds
with larger $\krho$ (Figure~\ref{f:SFE_krho}). Therefore, disruption
is quite insensitive to the degree of density concentration. Finally,
we explore the effect of trapped IR radiation using a usual spherical
symmetry prescription in which the associated radiation force is taken
proportional to the IR optical depth through a whole cloud (see
Figure~\ref{f:SFE_IR}). The minimum efficiency is lowered only
slightly for a  dust opacity of $\kappaIR \approx 5 \cm^2
\gram^{-1}$; a substantial reduction in $\SFEm$ requires $\kappaIR
\gtrsim 15 \cm^2 \gram^{-1}$, appropriate for dust-enriched
environments.

\subsection{Discussion}\label{s:dis}

In this paper, we have calculated $\SFEm$ required for cloud
disruption. Our results strongly suggest that GMCs are able to end
their lives within a single internal crossing time after the formation
of a large \HII region with stellar mass $\SFEm \Mcl$. While this
$\SFEm$ is a minimum, it would also be a reasonable estimate for the
net star formation efficiency (at least in the idealized spherical
case) for the following reasons. Consider a cloud with $\SFE <
\SFEm$. Since it cannot be destroyed by shell expansion, it will
continue to form stars, and $\SFE$ will increase. However, once $\SFE$
reaches $\SFEm$, if the disruption is rapid compared to the free-fall
time, the efficiency would not increase much beyond this value.

Similarly to our study, \citetalias{MQT10} obtained $\SFE_{\rm GMC}$
for five GMCs in various galactic environments. However, there are
several notable differences between their and our expansion
models. First, \citetalias{MQT10} made a distinction between the star
formation efficiency of the GMC and that of cluster-forming gas at the
center, the latter of which was fixed to 50\%, assuming that a half of
the ``cluster gas'' turns into stars and the other half goes into the
initial shell mass. Second, the shell expansion in \citetalias{MQT10}
begins at a radius (typically a few parsecs) that follows the observed
mass-radius relations of star clusters, whereas our model starts from
the initial Str\"{o}mgren radius. Third, \citetalias{MQT10} included
an additional inward force, of order $\sim G\Mcl^2/\Rcl^2$, due to
turbulent pressure, which is not considered in our models. We omit
this term as turbulent pressure is dominated by the largest scales,
and is difficult to model; swept-up gas may add either inward or
outward momentum to the shell. The expanding shells in
\citetalias{MQT10} have a higher surface density and stronger inward
force initially than in our model, which is partly compensated by the
outward forces they include to represent protostellar outflows and
direct/dust-reprocessed radiation. Despite these differences, the
conclusions of \citetalias{MQT10} that the $\SFE_{\rm GMC}$ increases
with surface density and that the disruption takes place in about a
free-fall time of the cloud are all qualitatively consistent with our
results.

\citet{mur11} estimated star formation efficiency of Galactic star
forming complexes by matching a WMAP sample of luminous free-free
sources to host GMCs with typical mass $\sim 10^6 \Msun$ and typical
surface density $\sim 10^2 \Msun \pc^{-2}$, finding $\varepsilon_{\rm
  GMC} = 0.08$ on average, for the luminosity limited samples (see his
Figure 2). These observed star formation efficiencies are in the same
range as our minimum efficiencies for the destruction criterion
$\vsh(\Rcl) = \vbind$ shown in Figure~\ref{f:SFE}(a), suggesting that
these clouds may be in the process of disruption by expanding \HII
regions.  Indeed, most of these star forming complexes show evidence
for expanding, bubble-like morphologies in infrared and radio
recombination lines, in excess of turbulent motions \citep{rah10,
  lee12}. \citet{mur11} also compared inward and outward forces to
find $\Fout > \Fin$ and $F_{\rm rad,eff} > F_{\rm therm,eff}$,
suggesting that expansion is driven by radiation in many of their
samples. \citet{gar14} presented a catalog of GMCs in the inner
southern Galaxy and estimated star formation efficiency of those
associated with ultracompact \HII regions. In their study, stellar
mass was inferred from far-IR luminosity from IRAS point-like source
catalog assuming that far-IR luminosity traces OB stellar population
with an age of less than $100\Myr$. They found the average star
formation efficiency of 3\%. While these observations are roughly
consistent with our results, the data need to be interpreted with
caution, because of selection bias \citep{mur11}, neglect of extended
emission \citep{gar14}, difficulty in identifying the boundary of a
cloud that is being destroyed, and uncertainties involved in lifetime
of stellar tracers. Also, it is important to distinguish between net
star formation efficiency $\SFE$ and observational estimates of
``current'' star formation efficiency, the latter of which does not
allow for the gas inflow/outflow experienced in the past as well as
future star formation before cloud destruction
\citep[e.g.,][]{mat00,fel11}.

Whether dust-reprocessed radiation can be effective in dispersing
star-forming clouds or not has been actively debated (e.g.,
\citetalias{KM09, MQT10}, \citealt{ski15}). Because the mean free path
of infrared photons can often be comparable to the system size, the
non-local nature of radiation makes it difficult to obtain the
solution of the radiative transfer equation. It is only in recent
years that the usage of the simplified prescription for $\FIR$ has
been tested by numerical simulations. For example, \citet{kru12b,
  kru13} used a flux limited diffusion scheme to investigate
matter-radiation interaction in a radiation-supported dusty
atmosphere.  They found that the photon trapping efficiency can be
greatly reduced by radiation-induced Rayleigh-Taylor instabilities
that provide channels for photons to escape, resulting in an
anti-correlation between matter and radiation. Using a more advanced
(variable Eddington tensor) algorithm, \citet{dav14} revisited these
calculations and found a reduced anti-correlation, corresponding to
stronger matter-radiation coupling. More recently, \citet{ski15}
adopted the M1 closure relation to run simulations of turbulent GMC
disruption by reprocessed radiation feedback, self-consistently
including self-gravitating collapse to produce sources of
radiation. This work showed that the usual trapping factor based on
the dust optical depth overestimates the radiation momentum deposition
rate by a factor of $\sim 4$--$5$, in part due to matter-radiation
anticorrelation, and in part due to the cancellation of radiation
forces where sources are distributed rather than centrally
concentrated. \citet{ski15} also showed that reprocessed radiation is
able to limit collapse only when the opacity is large, $\kappaIR > 15
\cm^2 \gram^{-1}$.  Here, we also find that reprocessing only
significantly affects the minimum efficiency when $\kappaIR$ is large.

While in this paper we exclusively focus on the effects of \HII region
expansion, diverse feedback mechanisms with different degrees of
importance are believed to operate in star-forming environments
(\citetalias{MQT10}; \citealt{fal10, kru14, mat15}).  For example,
outflows and jets from protostars feed turbulent motions within
cluster-forming clumps, prevent global collapse, and reduce star
formation efficiency therein \citep[e.g.,][]{mat00, mat07,
  wan10}. They are a dominant source of momentum injection before
massive stars form, but are unlikely to be capable of destroying the
intermediate-mass and massive clumps \citep[e.g.,][]{mat02, fal10,
  nak14}. Supernova explosions are regarded as the most powerful
feedback mechanism for driving turbulence in the ISM \citep{jou06},
providing vertical pressure support against gravity and regulating the
star formation rates in galactic disks \citep[e.g.,][]{ost11,
  kim13}. Supernovae may also significantly impact or destroy
molecular clouds if they are not dispersed by other feedback processes
during the lifetime of massive cluster stars
\citep[e.g.,][]{hen14,gee14,wal14}. Indeed, observations of
mixed-morphology supernova remnants suggest that this is often the
case \citep[e.g.,][]{rho98}.

Traditionally, thermal pressure of hot gas created by shocked stellar
winds is thought to dominate expansion of ionized bubbles
\citep{cas75, wea77, koo92}. The pressure of hot gas trapped within
the \HII region would push the inner boundary of the \HII region
outward, indirectly doing mechanical work on the swept-up shell. A
semi-analytic spherical model by \citet{mar14} suggests that shocked
wind pressure is more important than radiation pressure in driving
\HII region expansion. However, observed $X$-ray luminosity is lower
than the theoretical prediction of hot confined gas, casting doubt on
the effectiveness of stellar winds in controlling the dynamics of gas
around star clusters \citep{har09,yeh12,ros14}. Estimates of wind
energy lost via various mechanisms suggest that leakage of hot gas
through holes in the shell and/or turbulent mixing at hot--cold
interfaces can explain the observed low luminosity \citep{ros14}. The
approximate calculation by \citetalias{KM09} shows that a shocked wind
brings only a modest increase in effective outward force on a porous
shell, with a wind trapping factor of order unity. The star formation
efficiencies we present here are the minimum for disruption solely by
dynamical expansion of an \HII region, but other feedback process
could reduce $\SFEm$.

We now comment on caveats of the present study in various aspects.
First, we have assumed that the ionizing luminosity remains constant
during the shell expansion. For most clouds, this is acceptable as the
expansion time $t(\Rcl)$ is shorter than $3.8 \Myr$, a typical
main-sequence lifetime of ionizing stars \citep{mck97,kru06}. For
clouds located in the lower right corner in the $\Mcl$--$\Sigmacl$
plane (Figure~\ref{f:tR}), however, the ionizing output may experience
a significant drop before cloud disruption. With a decrease in the
outward force, the shell expansion would be slowed down and possibly
fall back due to gravity, implying that higher $\SFEm$ is required
than in our models. Second, while we assume spherical symmetry,
expanding shells may be subject to various non-radial instabilities
such as ionization front instability \citep{van62,kim14},
Rayleigh-Taylor instability, and Vishniac instability
\citep{gar96,wha08} in the early phase, and gravitational instability
assisted by external pressure in the late phase
\citep{wun10,iwa11,kim12}. At the nonlinear stage, the shell may break
up into pieces and create holes through which photons and ionized gas
can leak out, reducing feedback efficiency. Moreover, any shell that
forms in an inhomogeneous cloud would itself be inhomogeneous, such
that acceleration would be nonuniform (see below).

Third, our model considers a situation where the shell expansion is
driven solely by a single embedded \HII region, as in \citet{fal10}
and \citetalias{MQT10}. In reality, however, star formation in a
molecular cloud may be distributed spatially with a population of
subclusters \citep{mck97}. This results in expanding \HII regions of
various sizes interacting with each other. In early stage of
expansion, the momentum injection by individual subclusters may not
simply add up to the total due to cancelation; this effect is evident
in the simulations of \citet{ski15}. But once a main shell created by
the most luminous \HII region expands to a volume large compared to
that of most ionizing sources (assuming some degree of subcluster
concentration within the cloud), the effective radiation source is the
same as for a single central cluster.

More importantly, we have ignored blister-type \HII regions that can
effectively vent ionized gas through low density regions. Since real
turbulent clouds have a log-normal density distribution characterized
by many clumps and holes, even an initially fully embedded \HII region
is likely to transform into blister-type.  The analytic model of
\citet{mat02} in which all \HII regions are taken to be blister-type
showed that photoevaporative mass loss from small subclusters
(dominated by those around turnover in the cluster luminosity function
(\citealt{mck97})) alone can limit star formation efficiency of
galactic GMCs to below $\sim 10\%$, with the photodestruction time
scale decreasing as a function of the cloud mass. When both mass loss
mechanisms are considered, dynamical disruption by shell expansion is
more frequent than photodestruction alone, but significant
photoevaporative mass loss also occurs prior to disruption if clouds
survive for several free-fall times \citep{kru06}. However, these
analyses of photoevaporation did not allow for the reduced surface
area that gas confined in dense filaments presents to radiation, which
tends to reduce photoevaporation.  Indeed, \citet{dal12,dal13a} found
that less than 10\% of the gas is photoionized in their simulations.
The reduced effective area of clumpy clouds also reduces radiation
forces \citep{tho16,ras15}.  The relative importance of the mass loss
mechanisms in inhomogeneous clouds is difficult to assess and would
require numerical simulations (but see \citet{mat15}).

In inhomogeneous turbulent clouds, gas dispersal would take place in a
gradual, rather than impulsive, fashion. Both analytic and numerical
investigations of radiation-only feedback models show that broad
distributions of surface density and/or radiation flux cause locally
low-column, super-Eddington gas parcels to be ejected from the system
early on, while higher surface density parcels are ejected only later,
when the total luminosity of stars has increased. (e.g.,
\citealt{ras15,tho16}). For example, the blue shaded region in
Figure~\ref{f:SFE_krho} shows the range of net star formation
efficiency of turbulent clouds with width of lognormal surface density
distribution in the range $0<\sigma_{\ln \Sigma}<1$, predicted from
the analytic model of \citet{ras15} (see their Equation (21)), which
considers just radiation pressure on dust from non-ionizing radiation.
Stellar mass grows until there is little gas above the critical
surface density, making $\SFE$ increase beyond what is expected from
uniform case (lower bound). Note also that the inclusion of gas
pressure in our model is responsible for the dramatic difference in
efficiency for low $\Sigmacl$ clouds.

Because the current models consider only uniform spherical shells of
gas rather than a broadened surface density probability distribution
function, we underestimate the luminosity and hence $\SFEm$ required
to eject the denser clumps in a cloud via direct radiation
forces. However, the direct photoevaporation enabled by star formation
close enough to the cloud periphery would also tend to lower
$\SFEm$. Additional star formation can also in principle be triggered
in shocked shells by the collect and collapse process
\citep[e.g.,][]{elm77, hos06, dal07, iwa11, dal13b}. To address these
complex issues, it is necessary to perform three-dimensional radiation
hydrodynamic simulations of star cluster formation in a turbulent
cloud with both ionizing and non-ionizing radiation.

\acknowledgements We are grateful to the referee, Chris Matzner, for
insightful comments that greatly improved the manuscript. We also wish
to thank Kengo Tomida who made his version of {\it Athena} in
spherical coordinates available to us for use. The work of J.-G.K. was
supported by the National Research Foundation of Korea (NRF) grant
funded by the Korean Government (NRF-2014-Fostering Core Leaders of
the Future Basic Science Program). The work of W.-T.K. was supported
by the National Research Foundation of Korea (NRF) grant,
No.~2008-0060544, funded by the Korea government (MSIP). The work of
E.~C.~O. was supported by the NSF under grant AST-1312006.
Computation was supported by the Supercomputing Center/Korea Institute
of Science and Technology Information, with supercomputing resources
including technical support (KSC-2015-C3-027).

\appendix

\section{Light-to-mass ratio of Star Clusters}\label{s:ltom}

The photon output produced per unit stellar mass can differ greatly
from cluster to cluster at the low-mass end owing to stochastic
fluctuations in the stellar populations \citep{das12}, and/or due to
the correlation between cluster mass and the maximum stellar mass
\citep{wei06}. To assess the likely magnitude of this effect, we use a
new version of the SLUG code \citep{kru15} to simulate spectral
properties of star clusters as a function of mass.  In its simplest
setup, SLUG can simulate coeval stellar populations of finite mass and
predict a full spectrum based on libraries of stellar evolutionary
tracks and stellar atmosphere models.  We use the IMF of
\citet{cha03}, Starburst99 setting for spectral synthesis (which uses
the stellar atmosphere model for OB stars by \citet{pau01}), and the
Geneva library for stellar evolutionary tracks.

\begin{figure*}
\epsscale{1.0}\plotone{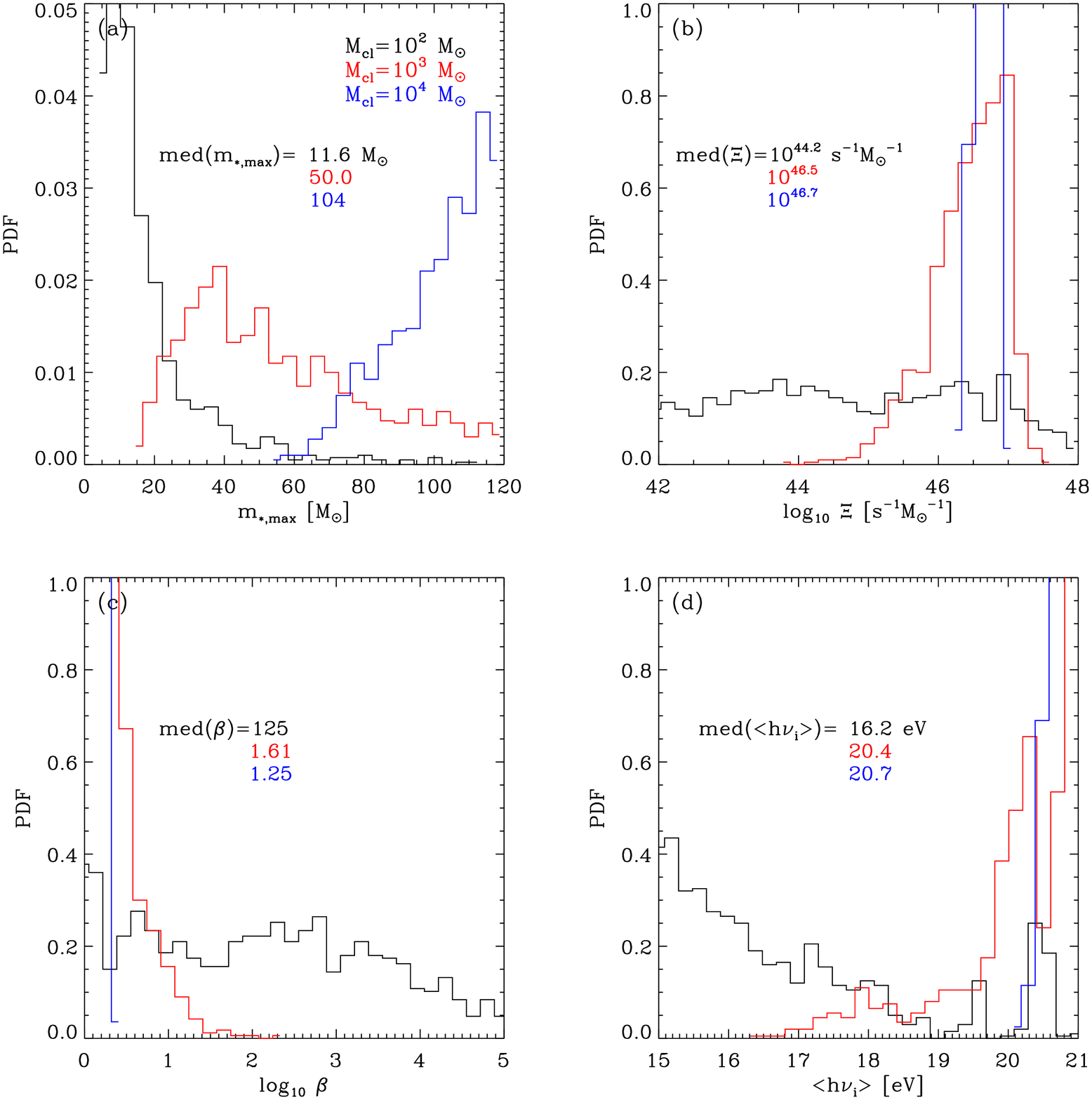} \caption{Probability distributions
  of (a) the maximum stellar mass $m_{*,\rm max}$, (b) the ionizing
  luminosity per stellar mass $\Qitom$, (c) the spectral parameter
  $\beta$, and (d) the mean ionizing photon energy $\hnui$ from $10^3$
  realizations of a star cluster with mass $\Mstar = 10^2\Msun$
  (black), $10^3\Msun$ (red), and $10^4\Msun$ (blue) using SLUG. The
  median values of the distributions are given in each
  panel.}\label{f:app1}
\end{figure*}

We ran 1000 simulations for different cluster mass bins logarithmically
spaced by 0.2 dex. Figure~\ref{f:app1} shows probability distributions
of the maximum stellar mass $m_{*,\rm max}$ in a cluster, the spectral
parameter $\beta$, the ionizing photon rate per stellar mass $\Qitom$,
and the mean ionizing photon energy $\hnui$ at $t=0.1 \Myr$ after
birth. The cases with the total cluster mass of $\Mstar = 10^2$,
$10^3$, and $10^4 M_{\odot}$ are shown as black, red, and blue
histograms, respectively. The median values of $m_{*,\rm  max}$,
$\beta$, $\Qitom$, and $\hnui$ are given in each panel. As expected,
$m_{*,\rm max}$ of massive clusters with $\Mstar=10^4 \Msun$ is close
to the theoretical maximum $120 \Msun$, whereas low-mass clusters with
$\Mstar=10^2 \Msun$ hardly contain O-type stars more massive than $\sim
20\Msun$. In the latter case, the distributions of $\beta$ and $\Qitom$
span several orders of magnitude due to high stochasticity.

\begin{figure*}
\epsscale{1.0}\plotone{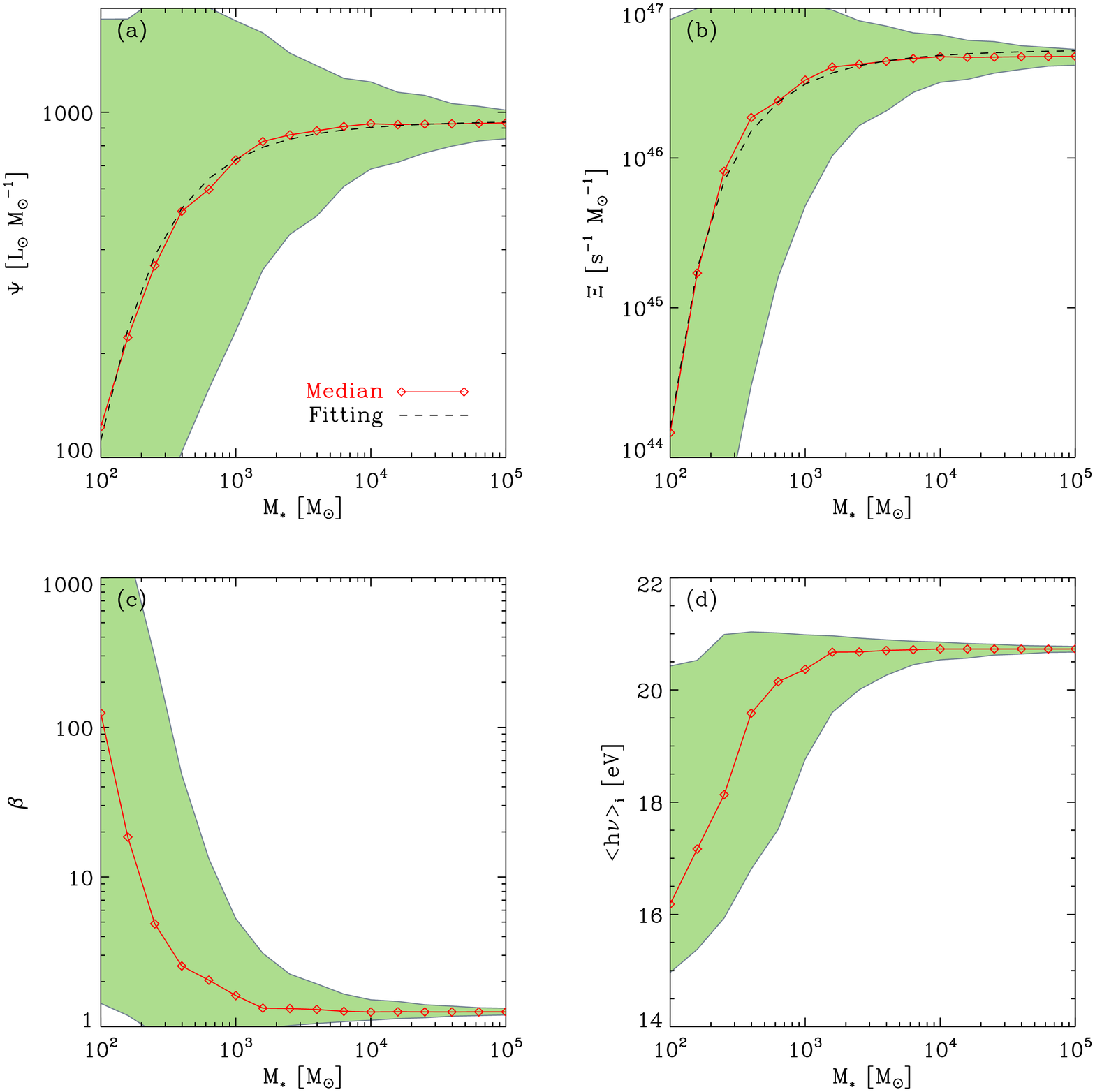} \caption{Dependence on the cluster
  mass of the median values of (a) the light-to-mass ratio $\Psi$, (b)
  the ionizing photon rate per unit mass $\Qitom$, (c) the spectral
  parameter $\beta$, and (d) the mean ionizing photon energy
  $\hnui$. The solid lines with diamonds are the median values, while
  the shaded regions represent the 10-90th percentile range from the
  Monte Carlo simulations. The dashed lines in the upper panels are
  our fits (Equations~\eqref{e:ltom} and \eqref{e:qitom}) to the
  median values of $\Psi$ and $\Qitom$. }\label{f:app2}
\end{figure*}

Figure~\ref{f:app2} plots the median values (red solid lines with
diamonds) and the 10--90th percentile range (green shade) of $\Ltom$,
$\Qitom$, $\beta$, and $\hnui$ as functions of the cluster mass. We
fit the median values to $\tilde{\Ltom}$ and $\tilde{\Qitom}$ using
\begin{equation}\label{e:ltom}
  \log_{10} \left( \frac{\tilde{\Ltom}}{\Lsun/\Msun} \right) =
  \dfrac{2.98 \mathcal{X}^6}{29.0 + \mathcal{X}^6}\,,
\end{equation}
and
\begin{equation}\label{e:qitom}
  \log_{10} \left( \frac{\tilde{\Qitom}}{1 \second^{-1}\Msun^{-1}}
  \right) = \dfrac{46.7 \mathcal{X}^6}{7.28 + \mathcal{X}^6}\,,
\end{equation}
where $\mathcal{X}=\log_{10} \Mstar/M_{\odot}$, which are drawn as
dashed lines in Figure~\ref{f:app2}.

For $\Mstar \gtrsim 10^4 \Msun$, $\tilde{\Ltom} = 943
\,L_{\odot}\,M_{\odot}^{-1}$ and $\tilde{\Qitom} = 5.05 \times 10^{46}
\second^{-1} \Msun^{-1}$, suitable for massive clusters that fully
sample the IMF \citep[see also][]{mur10}. But, both $\tilde{\Ltom}$
and $\tilde{\Qitom}$ decrease rapidly for clusters less massive than
$10^3 \Msun$. When $\Mstar = 10^2 \Msun$, for example,
$\tilde{\Qitom}$ is smaller by more than two orders of magnitude compared
to the fully sampled case. This is because a large number of samples
with small $\Mstar$ do not have (O-type) stars massive enough to emit
ionizing photons.

\section{Method of Numerical Simulations}\label{s:num}

Here we describe the method of direct numerical simulations for the
internal structure and expansion of an \HII region.  The equations of
hydrodynamics we solve in spherical symmetry are
\begin{subequations}
  \begin{equation}\label{e:gov1}
    \dfrac{\p \rho}{\p t} + \dfrac{\p}{\p r}(\rho v) + \dfrac{2\rho
      v}{r} = 0 \,,
  \end{equation}
  \begin{equation}\label{e:gov2}
    \dfrac{\p \rho v}{\p t} + \dfrac{\p}{\p r}(P + \rho v^2) +
    \dfrac{2\rho v^2}{r} = f_{\rm grav} + f_{\rm rad} \,,
  \end{equation}
  \begin{equation}\label{e:gov3}
    \dfrac{\p \rho_{\rm n}}{\p t} + \dfrac{\p}{\p r}(\rho_{\rm n} v) +
    \dfrac{2\rho_{\rm n} v}{r} = \muH(\mathcal{R} - \mathcal{I}) \,,
  \end{equation}
\end{subequations}
where $\rho$ is the total gas density, $\rho_{\rm n}$ is the neutral
gas density, $v$ is the radial velocity, and $P = (1 + x_e)n\kB T$ is
the gas pressure with the ionization fraction $x_e=1-\rho_n/\rho$. The
source terms in Equation \eqref{e:gov2} include the gravitational
acceleration ($f_{\rm grav} = -G M(<r) \rho/r^2$) and radiative force
(see Section~\ref{s:dr11}). In Equation \eqref{e:gov3},
$\mathcal{R}=\alphaB x_e^2n^2$ and $\mathcal{I}$ refer to the
recombination rate and ionization rate, respectively. We evolve
Equations \eqref{e:gov1}--\eqref{e:gov3} using a modified version of
the \textit{Athena} code in spherical coordinates \citep{sto08}.
Athena is an Eulerian code for magnetohydrodynamics based on a
directionally unsplit Godunov method. We use the van Leer algorithm
for time integration, a piecewise linear method for spatial
reconstruction, and the HLLC Riemann solver to compute the fluxes.

In order to handle radiation hydrodynamics coupled with
non-equilibrium chemistry, we implement a simple radiative transfer
algorithm based on the methods suggested by \citet{mel06b} and
\citet{kru07a}.  As in the \citetalias{Dr11} model, we regard the
radiation as being dichromatic, consisting of ionizing and
non-ionizing photons.  Let $\phi_{j-1/2}$ ($\psi_{j-1/2}$) and
$\phi_{j+1/2}$ ($\psi_{j+1/2}$) be the fractions of the ionizing
(non-ionizing) photons per second arriving at the inner and outer
boundaries of $j$-th cell, respectively.  The photon consumption rate
of ionizing radiation at the $j$-th shell is then computed as
$\Gamma_j = \Qi(\phi_{j-1/2} - \phi_{j+1/2})/\Delta V_j =
\Qi\phi_{j-1/2}(1 - e^{-\Delta\tau_j})/\Delta V_j$, where $\Delta V_j
= (4\pi/3)\left[(r_j + \Delta r/2)^3 - (r_j - \Delta r/2)^3\right]$ is
the shell volume and $\Delta \tau_j = (n_{{\rm n},j}\sigmapi +
n_j\sigmad)\Delta r$ is the optical depth across the $j$-th
cell. Here, $n_{{\rm n},j}$ and $n_j$ is the neutral and total density
of the $j$-th cell, respectively, and $\sigmapi = 6.3 \times 10^{-18}
\cm^2$ is the mean photoionization cross-section. The volumetric
photoionization rate of the $j$-th cell is then $\mathcal{I}_j =
\Gamma_j \times (n_{{\rm n},j}\sigmapi\Delta r)/\Delta\tau_j$.  The
calculation for non-ionizing radiation is carried out in a similar
way, with dust being the lone source of opacity.

We set the gas temperature according to the ionization fraction as
\begin{equation}
  T = \Tion - \left(\dfrac{1 - x_e}{1 + x_e}\right) (\Tion - T_{\rm n})\,,
\end{equation}
with $\Tion=10^4 \Kel$ and $T_{\rm n}=10^2 \Kel$ \citep{hen05}.
Clearly, $T=\Tion$ for $x_e=1$ and $T=T_{\rm n}$ for $x_e=0$. The
timestep for the radiation update is chosen by the requirement that
the relative changes in $T$ and $x_e$ should be less than 10\%. Since
this timestep is usually much shorter than that for the hydrodynamic
update, we subcycle the radiation update. The radiative force $f_{{\rm
    rad},j} = [\Li (\phi_{j - 1/2} - \phi_{j + 1/2}) + \Ln (\psi_{j -
    1/2} - \psi_{j + 1/2})]/(c\Delta V_j)$ at the $j$-th cell is added
to the momentum flux explicitly at the end of every subcycle.  To
ensure accuracy of hydrodynamics, we impose an additional constraint
on the hydrodynamic timestep such that the temperature and ionization
fraction should not change by more than a factor of 4 between the
hydrodynamic updates.

In all models, we fix the spatial resolution to $\Delta r = 0.005 \pc$
and place the inner boundary at $0.1 \pc$. The location of the outer
boundary is chosen large enough to cover the stalling radius of an
expanding shell. We adopt the outflow boundary conditions at both inner
and outer radial boundaries of the simulation domain.  We have tested
our implementation of the ionization chemistry against a standard
problem for supersonic (R-type) propagation of an ionization front in
a dustless, static medium (e.g., Test 1 in \citealt{ili06}), confirming
that our numerical results reproduce the analytic solutions within
errors of 3\%. We then apply the code to our main problem, namely,
expansion of a dusty \HII region in a stratified cloud with $\krho=1$
and $T=T_{\rm n}$.  Without star formation and feedback, the cloud is
supposed to be in force balance between gravity and turbulent pressure.
Since our one-dimensional models are unable to handle turbulence
properly, we enforce the hydrostatic balance in the outer envelope
unaffected by shell expansion. This is done effectively by turning off
all body forces and taking $v=0$ outside the shock front. Section
\ref{s:sim} presents our numerical results for models with fixed
$\beta=1.5$ and $\gamma=11.1$, and varying density and $\Qi$.

\hspace{1cm}



\begin{thebibliography}
\expandafter\ifx\csname natexlab\endcsname\relax\def\natexlab#1{#1}\fi

\bibitem[{{Andr{\'e}} {et~al.}(2014){Andr{\'e}}, {Di Francesco},
  {Ward-Thompson}, {Inutsuka}, {Pudritz}, \& {Pineda}}]{and14}
{Andr{\'e}}, P., {Di Francesco}, J., {Ward-Thompson}, D., {Inutsuka}, S.-I.,
  {Pudritz}, R.~E., \& {Pineda}, J.~E. 2014, Protostars and Planets VI, 27

\bibitem[{{Arthur} {et~al.}(2011){Arthur}, {Henney}, {Mellema}, {de Colle}, \&
  {V{\'a}zquez-Semadeni}}]{art11}
{Arthur}, S.~J., {Henney}, W.~J., {Mellema}, G., {de Colle}, F., \&
  {V{\'a}zquez-Semadeni}, E. 2011, \mnras, 414, 1747

\bibitem[{{Arthur} {et~al.}(2004){Arthur}, {Kurtz}, {Franco}, \&
  {Albarr{\'a}n}}]{art04}
{Arthur}, S.~J., {Kurtz}, S.~E., {Franco}, J., \& {Albarr{\'a}n}, M.~Y. 2004,
  \apj, 608, 282

\bibitem[{{Ashman} \& {Zepf}(2001)}]{ash01}
{Ashman}, K.~M., \& {Zepf}, S.~E. 2001, \aj, 122, 1888

\bibitem[Banerjee \& Kroupa(2015)]{ban15} Banerjee, S., \& Kroupa,
  P.\ 2015, arXiv:1512.03074

\bibitem[{{Bertoldi} \& {McKee}(1992)}]{ber92}
{Bertoldi}, F., \& {McKee}, C.~F. 1992, \apj, 395, 140

\bibitem[{{Carpenter}(2000)}]{car00} {Carpenter}, J.~M. 2000, \aj,
  120, 3139

\bibitem[{{Castor} {et~al.}(1975){Castor}, {McCray}, \& {Weaver}}]{cas75}
{Castor}, J., {McCray}, R., \& {Weaver}, R. 1975, \apjl, 200, L107

\bibitem[{{Chabrier}(2003)}]{cha03} {Chabrier}, G. 2003, \pasp, 115,
  763

\bibitem[{{Da Rio} {et~al.}(2014){Da Rio}, {Tan}, \& {Jaehnig}}]{dar14}
{Da Rio}, N., {Tan}, J.~C., \& {Jaehnig}, K. 2014, \apj, 795, 55

\bibitem[{{da Silva} {et~al.}(2012){da Silva}, {Fumagalli}, \&
  {Krumholz}}]{das12}
{da Silva}, R.~L., {Fumagalli}, M., \& {Krumholz}, M. 2012, \apj, 745, 145

\bibitem[{{Dale} {et~al.}(2007){Dale}, {Bonnell}, \& {Whitworth}}]{dal07}
{Dale}, J.~E., {Bonnell}, I.~A., \& {Whitworth}, A.~P. 2007, \mnras, 375, 1291

\bibitem[{{Dale} {et~al.}(2012){Dale}, {Ercolano}, \& {Bonnell}}]{dal12}
{Dale}, J.~E., {Ercolano}, B., \& {Bonnell}, I.~A. 2012, \mnras, 424, 377

\bibitem[{{Dale} {et~al.}(2013){Dale}, {Ercolano}, \&
    {Bonnell}}]{dal13a} ---. 2013, \mnras, 430, 234

\bibitem[{{Dale} {et~al.}(2013){Dale}, {Ercolano}, \&
    {Bonnell}}]{dal13b} ---. 2013, \mnras, 431, 1062

\bibitem[{{Davis} {et~al.}(2014){Davis}, {Jiang}, {Stone}, \& {Murray}}]{dav14}
{Davis}, S.~W., {Jiang}, Y.-F., {Stone}, J.~M., \& {Murray}, N. 2014, \apj,
  796, 107

\bibitem[{{Dobbs} {et~al.}(2014){Dobbs}, {Krumholz}, {Ballesteros-Paredes},
  {Bolatto}, {Fukui}, {Heyer}, {Low}, {Ostriker}, \&
  {V{\'a}zquez-Semadeni}}]{dob14}
{Dobbs}, C.~L., {et~al.} 2014, Protostars and Planets VI, 3

\bibitem[{{Draine}(2011)}]{Dr11}
{Draine}, B.~T. 2011, \apj, 732, 100 (Dr11)

\bibitem[Draine(2011)]{dr11} Draine, B.~T.\ 2011,
  Physics of the Interstellar and Intergalactic Medium by Bruce
  T.~Draine.~Princeton University Press, 2011

\bibitem[{{Elmegreen}(1983)}]{elm83}
{Elmegreen}, B.~G. 1983, \mnras, 203, 1011

\bibitem[{{Elmegreen} \& {Efremov}(1997)}]{elm97}
{Elmegreen}, B.~G., \& {Efremov}, Y.~N. 1997, \apj, 480, 235

\bibitem[{{Elmegreen} \& {Lada}(1977)}]{elm77}
{Elmegreen}, B.~G., \& {Lada}, C.~J. 1977, \apj, 214, 725

\bibitem[{{Elmegreen} \& {Scalo}(2004)}]{elm04}
{Elmegreen}, B.~G., \& {Scalo}, J. 2004, \araa, 42, 211

\bibitem[{{Evans} {et~al.}(2009){Evans}, {Dunham}, {J{\o}rgensen}, {Enoch},
  {Mer{\'{\i}}n}, {van Dishoeck}, {Alcal{\'a}}, {Myers}, {Stapelfeldt},
  {Huard}, {Allen}, {Harvey}, {van Kempen}, {Blake}, {Koerner}, {Mundy},
  {Padgett}, \& {Sargent}}]{eva09}
{Evans}, II, N.~J., {et~al.} 2009, \apjs, 181, 321

\bibitem[{{Fall} {et~al.}(2010){Fall}, {Krumholz}, \& {Matzner}}]{fal10}
{Fall}, S.~M., {Krumholz}, M.~R., \& {Matzner}, C.~D. 2010, \apjl, 710, L142

\bibitem[Feldmann \& Gnedin(2011)]{fel11} Feldmann, R., \& Gnedin,
  N.~Y.\ 2011, \apjl, 727, L12

\bibitem[Franco et al.(1990)]{fra90} Franco, J.,
  Tenorio-Tagle, G., \& Bodenheimer, P.\ 1990, \apj, 349, 126

\bibitem[{{Franco} {et~al.}(1994){Franco}, {Shore}, \& {Tenorio-Tagle}}]{fra94}
{Franco}, J., {Shore}, S.~N., \& {Tenorio-Tagle}, G. 1994, \apj, 436, 795

\bibitem[Garc{\'{\i}}a et al.(2014)]{gar14}
  Garc{\'{\i}}a, P., Bronfman, L., Nyman, L.-{\AA}., Dame, T.~M., \&
  Luna, A.\ 2014, \apjs, 212, 2

\bibitem[Garcia-Segura \& Franco(1996)]{gar96}
  Garcia-Segura, G., \& Franco, J.\ 1996, \apj, 469, 171

\bibitem[Geen et al.(2015)]{gee14} Geen, S., Rosdahl, J., Blaizot, J.,
  Devriendt, J., \& Slyz, A.\ 2015, \mnras, 448, 3248

\bibitem[Geen et al.(2015)]{gee15} Geen, S., Hennebelle,
  P., Tremblin, P., \& Rosdahl, J.\ 2015, \mnras, 454, 4484

\bibitem[{{Genzel} {et~al.}(2010){Genzel}, {Tacconi}, {Gracia-Carpio},
  {Sternberg}, {Cooper}, {Shapiro}, {Bolatto}, {Bouch{\'e}}, {Bournaud},
  {Burkert}, {Combes}, {Comerford}, {Cox}, {Davis}, {Schreiber},
  {Garcia-Burillo}, {Lutz}, {Naab}, {Neri}, {Omont}, {Shapley}, \&
  {Weiner}}]{gen10}
{Genzel}, R., {et~al.} 2010, \mnras, 407, 2091

\bibitem[{{Goldbaum} {et~al.}(2011){Goldbaum}, {Krumholz}, {Matzner}, \&
  {McKee}}]{gol11}
{Goldbaum}, N.~J., {Krumholz}, M.~R., {Matzner}, C.~D., \& {McKee}, C.~F. 2011,
  \apj, 738, 101

\bibitem[{{Goodwin}(1997)}]{goo97}
{Goodwin}, S.~P. 1997, \mnras, 284, 785

\bibitem[{{Harper-Clark}(2011)}]{har11}
{Harper-Clark}, E. 2011, PhD thesis, University of Toronto (Canada)

\bibitem[{{Harper-Clark} \& {Murray}(2009)}]{har09}
{Harper-Clark}, E., \& {Murray}, N. 2009, \apj, 693, 1696

\bibitem[{{Heiles} \& {Troland}(2003)}]{hei03}
{Heiles}, C., \& {Troland}, T.~H. 2003, \apj, 586, 1067

\bibitem[Hennebelle \& Iffrig(2014)]{hen14} Hennebelle, P., \& Iffrig,
  O.\ 2014, \aap, 570, A81

\bibitem[Henney \& Arthur(1998)]{hen98} Henney, W.~J., \& Arthur,
  S.~J.\ 1998, \aj, 116, 322

\bibitem[{{Henney} {et~al.}(2005){Henney}, {Arthur}, {Williams}, \&
  {Ferland}}]{hen05}
{Henney}, W.~J., {Arthur}, S.~J., {Williams}, R.~J.~R., \& {Ferland}, G.~J.
  2005, \apj, 621, 328

\bibitem[{{Hillenbrand} \& {Hartmann}(1998)}]{hil98}
{Hillenbrand}, L.~A., \& {Hartmann}, L.~W. 1998, \apj, 492, 540

\bibitem[{{Hills}(1980)}]{hil80}
{Hills}, J.~G. 1980, \apj, 235, 986

\bibitem[{{Hopkins} {et~al.}(2011){Hopkins}, {Quataert}, \& {Murray}}]{hop11}
{Hopkins}, P.~F., {Quataert}, E., \& {Murray}, N. 2011, \mnras, 417, 950

\bibitem[{{Hosokawa} \& {Inutsuka}(2006)}]{hos06}
{Hosokawa}, T., \& {Inutsuka}, S.-i. 2006, \apj, 646, 240

\bibitem[{{Iliev} {et~al.}(2006){Iliev}, {Ciardi}, {Alvarez}, {Maselli},
  {Ferrara}, {Gnedin}, {Mellema}, {Nakamoto}, {Norman}, {Razoumov},
  {Rijkhorst}, {Ritzerveld}, {Shapiro}, {Susa}, {Umemura}, \& {Whalen}}]{ili06}
{Iliev}, I.~T., {et~al.} 2006, \mnras, 371, 1057

\bibitem[{{Iwasaki} {et~al.}(2011){Iwasaki}, {Inutsuka}, \&
    {Tsuribe}}]{iwa11} {Iwasaki}, K., {Inutsuka}, S.-i., \& {Tsuribe},
  T. 2011, \apj, 733, 16

\bibitem[{{Joung} \& {Mac Low}(2006)}]{jou06}
{Joung}, M.~K.~R., \& {Mac Low}, M.-M. 2006, \apj, 653, 1266

\bibitem[Kennicutt \& Evans(2012)]{ken12} Kennicutt, R.~C., \& Evans,
  N.~J.\ 2012, \araa, 50, 531

\bibitem[Keto(2002)]{ket02} Keto, E.\ 2002, \apj, 580,
  980

\bibitem[{{Keto}(2003)}]{ket03}
{Keto}, E. 2003, \apj, 599, 1196

\bibitem[{{Keto}(2007)}]{ket07}
---. 2007, \apj, 666, 976

\bibitem[{{Kim} {et~al.}(2013){Kim}, {Ostriker}, \& {Kim}}]{kim13}
{Kim}, C.-G., {Ostriker}, E.~C., \& {Kim}, W.-T. 2013, \apj, 776, 1

\bibitem[Kim et al.(2012)]{kim12} Kim, J.-G., Kim, W.-T., Seo, Y.~M.,
  \& Hong, S.~S.\ 2012, \apj, 761, 131

\bibitem[Kim \& Kim(2014)]{kim14} Kim, J.-G., \& Kim, W.-T.\ 2014,
  \apj, 797, 135

\bibitem[{{Koo} \& {McKee}(1992)}]{koo92}
{Koo}, B.-C., \& {McKee}, C.~F. 1992, \apj, 388, 93

\bibitem[{{Kroupa} \& {Boily}(2002)}]{kro02}
{Kroupa}, P., \& {Boily}, C.~M. 2002, \mnras, 336, 1188

\bibitem[{{Krumholz} {et~al.}(2012){Krumholz}, {Dekel}, \& {McKee}}]{kru12a}
{Krumholz}, M.~R., {Dekel}, A., \& {McKee}, C.~F. 2012, \apj, 745, 69

\bibitem[Krumholz et al.(2015)]{kru15} Krumholz, M.~R.,
  Fumagalli, M., da Silva, R.~L., Rendahl, T., \& Parra, J.\ 2015,
  \mnras, 452, 1447

\bibitem[{{Krumholz} \& {Matzner}(2009)}]{KM09}
{Krumholz}, M.~R., \& {Matzner}, C.~D. 2009, \apj, 703, 1352 (KM09)

\bibitem[{{Krumholz} {et~al.}(2006){Krumholz}, {Matzner}, \& {McKee}}]{kru06}
{Krumholz}, M.~R., {Matzner}, C.~D., \& {McKee}, C.~F. 2006, \apj, 653, 361

\bibitem[{{Krumholz} \& {McKee}(2005)}]{kru05}
{Krumholz}, M.~R., \& {McKee}, C.~F. 2005, \apj, 630, 250

\bibitem[{{Krumholz} {et~al.}(2007){Krumholz}, {Stone}, \& {Gardiner}}]{kru07a}
{Krumholz}, M.~R., {Stone}, J.~M., \& {Gardiner}, T.~A. 2007, \apj, 671, 518

\bibitem[{{Krumholz} \& {Tan}(2007)}]{kru07b}
{Krumholz}, M.~R., \& {Tan}, J.~C. 2007, \apj, 654, 304

\bibitem[{{Krumholz} \& {Thompson}(2012)}]{kru12b}
{Krumholz}, M.~R., \& {Thompson}, T.~A. 2012, \apj, 760, 155

\bibitem[{{Krumholz} \& {Thompson}(2013)}]{kru13}
---. 2013, \mnras, 434, 2329

\bibitem[{{Krumholz} {et~al.}(2014){Krumholz}, {Bate}, {Arce}, {Dale},
  {Gutermuth}, {Klein}, {Li}, {Nakamura}, \& {Zhang}}]{kru14}
{Krumholz}, M.~R., {et~al.} 2014, Protostars and Planets VI, 243

\bibitem[{{Lada} \& {Lada}(2003)}]{lad03}
{Lada}, C.~J., \& {Lada}, E.~A. 2003, \araa, 41, 57

\bibitem[{{Lee} {et~al.}(2012){Lee}, {Murray}, \& {Rahman}}]{lee12}
{Lee}, E.~J., {Murray}, N., \& {Rahman}, M. 2012, \apj, 752, 146

\bibitem[{{Leroy} {et~al.}(2013){Leroy}, {Walter}, {Sandstrom}, {Schruba},
  {Munoz-Mateos}, {Bigiel}, {Bolatto}, {Brinks}, {de Blok}, {Meidt}, {Rix},
  {Rosolowsky}, {Schinnerer}, {Schuster}, \& {Usero}}]{ler13}
{Leroy}, A.~K., {et~al.} 2013, \aj, 146, 19

\bibitem[{{Lopez} {et~al.}(2011){Lopez}, {Krumholz}, {Bolatto}, {Prochaska}, \&
  {Ramirez-Ruiz}}]{lop11}
{Lopez}, L.~A., {Krumholz}, M.~R., {Bolatto}, A.~D., {Prochaska}, J.~X., \&
  {Ramirez-Ruiz}, E. 2011, \apj, 731, 91

\bibitem[{{Lopez} {et~al.}(2014){Lopez}, {Krumholz}, {Bolatto}, {Prochaska},
  {Ramirez-Ruiz}, \& {Castro}}]{lop14}
{Lopez}, L.~A., {Krumholz}, M.~R., {Bolatto}, A.~D., {Prochaska}, J.~X.,
  {Ramirez-Ruiz}, E., \& {Castro}, D. 2014, \apj, 795, 121

\bibitem[Mart{\'{\i}}nez-Gonz{\'a}lez et al.(2014)]{mar14}
  Mart{\'{\i}}nez-Gonz{\'a}lez, S., Silich, S., \& Tenorio-Tagle,
  G.\ 2014, \apj, 785, 164

\bibitem[{{Mathews}(1967)}]{mat67}
{Mathews}, W.~G. 1967, \apj, 147, 965

\bibitem[{{Matzner}(2002)}]{mat02}
{Matzner}, C.~D. 2002, \apj, 566, 302

\bibitem[{{Matzner}(2007)}]{mat07}
---. 2007, \apj, 659, 1394

\bibitem[Matzner \& Jumper(2015)]{mat15} Matzner, C.~D.,
  \& Jumper, P.~H.\ 2015, arXiv:1511.03269

\bibitem[{{Matzner} \& {McKee}(2000)}]{mat00}
{Matzner}, C.~D., \& {McKee}, C.~F. 2000, \apj, 545, 364

\bibitem[{{McKee} \& {Ostriker}(2007)}]{mck07}
{McKee}, C.~F., \& {Ostriker}, E.~C. 2007, \araa, 45, 565

\bibitem[McKee \& Williams(1997)]{mck97} McKee, C.~F.,
  \& Williams, J.~P.\ 1997, \apj, 476, 144

\bibitem[{{Meier} {et~al.}(2002){Meier}, {Turner}, \& {Beck}}]{mei02}
{Meier}, D.~S., {Turner}, J.~L., \& {Beck}, S.~C. 2002, \aj, 124, 877

\bibitem[{{Mellema} {et~al.}(2006){Mellema}, {Iliev}, {Alvarez}, \&
  {Shapiro}}]{mel06b}
{Mellema}, G., {Iliev}, I.~T., {Alvarez}, M.~A., \& {Shapiro}, P.~R. 2006, \na,
  11, 374

\bibitem[{{Murray}(2011)}]{mur11}
{Murray}, N. 2011, \apj, 729, 133

\bibitem[Murray et al.(2010)]{MQT10} Murray, N., Quataert, E., \&
  Thompson, T.~A.\ 2010, \apj, 709, 191 (MQT10)

\bibitem[{{Murray} \& {Rahman}(2010)}]{mur10}
{Murray}, N., \& {Rahman}, M. 2010, \apj, 709, 424

\bibitem[{{Myers} {et~al.}(1986){Myers}, {Dame}, {Thaddeus}, {Cohen},
  {Silverberg}, {Dwek}, \& {Hauser}}]{mye86}
{Myers}, P.~C., {Dame}, T.~M., {Thaddeus}, P., {Cohen}, R.~S., {Silverberg},
  R.~F., {Dwek}, E., \& {Hauser}, M.~G. 1986, \apj, 301, 398

\bibitem[{{Nakamura} \& {Li}(2014)}]{nak14}
{Nakamura}, F., \& {Li}, Z.-Y. 2014, \apj, 783, 115

\bibitem[Osterbrock(1989)]{ost89} Osterbrock, D.~E.\ 1989,
  Astrophysics of Gaseous Nebulae and Active Galactic Nuclei (Mill
  Valley, CA, University Science Books)

\bibitem[{{Ostriker} \& {Shetty}(2011)}]{ost11}
{Ostriker}, E.~C., \& {Shetty}, R. 2011, \apj, 731, 41

\bibitem[{{Pauldrach} {et~al.}(2001){Pauldrach}, {Hoffmann}, \&
  {Lennon}}]{pau01}
{Pauldrach}, A.~W.~A., {Hoffmann}, T.~L., \& {Lennon}, M. 2001, \aap, 375, 161

\bibitem[{{Petrosian} {et~al.}(1972){Petrosian}, {Silk}, \& {Field}}]{pet72}
{Petrosian}, V., {Silk}, J., \& {Field}, G.~B. 1972, \apjl, 177, L69

\bibitem[{{Rahman} \& {Murray}(2010)}]{rah10}
{Rahman}, M., \& {Murray}, N. 2010, \apj, 719, 1104

\bibitem[{{Raskutti} {et~al.}(2015){Raskutti}, {Ostriker}, \&
    {Skinner}}]{ras15} {Raskutti}, S., {Ostriker}, E.~C., \&
  {Skinner}, M.~A. ApJ, submitted

\bibitem[Rho \& Petre(1998)]{rho98} Rho, J., \& Petre, R.\ 1998,
  \apjl, 503, L167

\bibitem[{{Rosen} {et~al.}(2014){Rosen}, {Lopez}, {Krumholz}, \&
  {Ramirez-Ruiz}}]{ros14}
{Rosen}, A.~L., {Lopez}, L.~A., {Krumholz}, M.~R., \& {Ramirez-Ruiz}, E. 2014,
  \mnras, 442, 2701

\bibitem[{{Scoville} {et~al.}(2001){Scoville}, {Polletta}, {Ewald}, {Stolovy},
  {Thompson}, \& {Rieke}}]{sco01}
{Scoville}, N.~Z., {Polletta}, M., {Ewald}, S., {Stolovy}, S.~R., {Thompson},
  R., \& {Rieke}, M. 2001, \aj, 122, 3017

\bibitem[{{Semenov} {et~al.}(2003){Semenov}, {Henning}, {Helling}, {Ilgner}, \&
  {Sedlmayr}}]{sem03}
{Semenov}, D., {Henning}, T., {Helling}, C., {Ilgner}, M., \& {Sedlmayr}, E.
  2003, \aap, 410, 611

\bibitem[Shu(1992)]{shu92} Shu, F.~H.\ 1992, The physics of
  astrophysics.~Volume II: Gas dynamics., by Shu, F.~H..~ University
  Science Books, Mill Valley, CA (USA), 1992

\bibitem[Shu et al.(2002)]{shu02} Shu, F.~H., Lizano, S., Galli, D.,
  Cant{\'o}, J., \& Laughlin, G.\ 2002, \apj, 580, 969

\bibitem[{{Skinner} \& {Ostriker}(2013)}]{ski13}
{Skinner}, M.~A., \& {Ostriker}, E.~C. 2013, \apjs, 206, 21

\bibitem[Skinner \& Ostriker(2015)]{ski15} Skinner,
  M.~A., \& Ostriker, E.~C.\ 2015, \apj, 809, 187

\bibitem[{{Spitzer}(1978)}]{spi78} {Spitzer}, L. 1978, {Physical
  processes in the interstellar medium}

\bibitem[{{Stone} {et~al.}(2008){Stone}, {Gardiner}, {Teuben},
    {Hawley}, \& {Simon}}]{sto08} {Stone}, J.~M., {Gardiner}, T.~A.,
  {Teuben}, P., {Hawley}, J.~F., \& {Simon}, J.~B. 2008, \apjs, 178,
  137

\bibitem[Tan et al.(2006)]{tan06} Tan, J.~C., Krumholz, M.~R., \&
  McKee, C.~F.\ 2006, \apjl, 641, L121

\bibitem[{{Tan} {et~al.}(2014){Tan}, {Beltr{\'a}n}, {Caselli},
    {Fontani}, {Fuente}, {Krumholz}, {McKee}, \& {Stolte}}]{tan14}
  {Tan}, J.~C., {Beltr{\'a}n}, M.~T., {Caselli}, P., {Fontani}, F.,
  {Fuente}, A., {Krumholz}, M.~R., {McKee}, C.~F., \& {Stolte},
  A. 2014, Protostars and Planets VI, 149

\bibitem[{{Thompson} {et~al.}(2015){Thompson}, {Fabian}, {Quataert},
    \& {Murray}}]{tho15} {Thompson}, T.~A., {Fabian}, A.~C.,
  {Quataert}, E., \& {Murray}, N. 2015, \mnras, 449, 147

\bibitem[Thompson \& Krumholz(2016)]{tho16} Thompson, T.~A., \&
  Krumholz, M.~R.\ 2016, \mnras, 455, 334

\bibitem[Tremblin et al.(2014)]{tre14} Tremblin, P., Anderson, L.~D.,
  Didelon, P., et al.\ 2014, \aap, 568, A4

\bibitem[{{Turner} {et~al.}(2015){Turner}, {Beck}, {Benford},
    {Consiglio}, {Ho}, {Kov{\'a}cs}, {Meier}, \& {Zhao}}]{tur15}
  {Turner}, J.~L., {Beck}, S.~C., {Benford}, D.~J., {Consiglio},
  S.~M., {Ho}, P.~T.~P., {Kov{\'a}cs}, A., {Meier}, D.~S., \& {Zhao},
  J.-H. 2015, \nat, 519, 331

\bibitem[Vandervoort(1962)]{van62} Vandervoort, P.~O.\ 1962, \apj,
  135, 212

\bibitem[Walch \& Naab(2015)]{wal14} Walch, S., \& Naab, T.\ 2015,
  \mnras, 451, 2757

\bibitem[{{Wang} {et~al.}(2010){Wang}, {Li}, {Abel}, \& {Nakamura}}]{wan10}
{Wang}, P., {Li}, Z.-Y., {Abel}, T., \& {Nakamura}, F. 2010, \apj, 709, 27

\bibitem[{{Weaver} {et~al.}(1977){Weaver}, {McCray}, {Castor},
    {Shapiro}, \& {Moore}}]{wea77} {Weaver}, R., {McCray}, R.,
  {Castor}, J., {Shapiro}, P., \& {Moore}, R. 1977, \apj, 218, 377

\bibitem[{{Weidner} \& {Kroupa}(2006)}]{wei06} {Weidner}, C., \&
  {Kroupa}, P. 2006, \mnras, 365, 1333

\bibitem[Whalen \& Norman(2008)]{wha08} Whalen, D.~J., \& Norman,
  M.~L.\ 2008, \apj, 672, 287

\bibitem[{{Whitworth}(1979)}]{whi79} {Whitworth}, A. 1979, \mnras,
  186, 59

\bibitem[{{Williams} \& {McKee}(1997)}]{wil97} {Williams}, J.~P., \&
  {McKee}, C.~F. 1997, \apj, 476, 166

\bibitem[W{\"u}nsch et al.(2010)]{wun10} W{\"u}nsch, R., Dale, J.~E.,
  Palou{s}, J., \& Whitworth, A.~P.\ 2010, \mnras, 407, 1963

\bibitem[{{Yeh} \& {Matzner}(2012)}]{yeh12} {Yeh}, S.~C.~C., \&
  {Matzner}, C.~D. 2012, \apj, 757, 108

\end{thebibliography}

\end{document}